\documentclass[%
 reprint,
 amsmath,
 amssymb,
 aps,
 longbibliography,
]{revtex4-1}

\usepackage{graphicx}
\usepackage{dcolumn}
\usepackage{bm}
\usepackage{hyperref}
\usepackage{braket}

\newcommand*{\affmark}[1][*]{\textsuperscript{#1}}

\begin{document}


\title{Non-reciprocal quantum Hall devices with driven edge magnetoplasmons in 2-dimensional materials}

\author{S. Bosco\affmark[1,3]}
\email {bosco@physik.rwth-aachen.de}
\author{D. P. DiVincenzo\affmark[1,2,3]}
 \email{d.divincenzo@fz-juelich.de}
\affiliation{
\affmark[1]Institute for Quantum Information, RWTH Aachen University,                                
  D-52056
  Aachen,                              
  Germany
}

\affiliation{
  \affmark[2]Peter Gr\"{u}nberg Institute, Theoretical Nanoelectronics,
    Forschungszentrum J\"{u}lich,
  D-52425
  J\"{u}lich,
  Germany
}

\affiliation{
\affmark[3]J\"{u}lich-Aachen Research Alliance (JARA),
    Fundamentals of Future Information Technologies,
  D-52425
  J\"{u}lich,
  Germany
}

\date{\today}

\begin{abstract}

We develop a theory that describes the response of non-reciprocal devices employing 2-dimensional materials in the quantum Hall regime capacitively coupled to external electrodes.
As the conduction in these devices is understood to be associated to the edge magnetoplasmons (EMPs), we first investigate the EMP problem by using the linear response theory in the random phase approximation.
Our model can incorporate several cases, that were often treated on different grounds in literature. 
In particular, we analyze plasmonic excitations supported by smooth and sharp confining potential in 2-dimensional electron gas, and in monolayer graphene, and we point out the similarities and differences in these materials.
We also account for a general time-dependent external drive applied to the system.
Finally, we describe the behavior of a non-reciprocal quantum Hall device: the response contains additional resonant features, which were not foreseen from previous models.


\end{abstract}

\pacs{Valid PACS appear here}
\maketitle


\section{\label{sec:intro}Introduction}

Non-reciprocal devices, such as gyrators and circulators, are key components for modern microwave engineering.
They allow a variety of operations required for several applications, including qubit control and thermal noise reduction.

An ideal gyrator induces a $\pi$-shift between signals moving in opposite directions: this behavior is captured by the scattering ($S$) matrix \cite{Pozar}
\begin{equation}
S= e^{i\theta}
\left(
\begin{array}{cc}
 0 & -1 \\
 1 & 0 \\
\end{array}
\right).
\label{eq:s-g-matrix}
\end{equation}

An implementation for these devices that guarantees good miniaturization was recently proposed by Viola and DiVincenzo (VD) \cite{Viola-DiVincenzo}.
The main idea is to use a 2-dimensional conductor in the quantum Hall (QH) regime \cite{QuantumHallGirvin} capacitively coupled to external metal electrodes.
The voltage applied to the electrodes excites the magneto plasmons at the edge of the conductor (EMP): as they move chirally, with direction dependent on the sign of the applied magnetic field, they are responsible for the non-reciprocal behavior of the device.

Further developments both on the theoretical \cite{Viola-DiVincenzo, Bosco, Placke} and on the experimental side \cite{Reilly}, showed that the VD model is very useful to understand the main behavior of these devices, but some questions were left unanswered.
For example, it is well-known \cite{Glazman, Glazman2,  JohnsonVignale, Mikhailov} that the edge of 2-dimensional conductors supports several plasmonic modes, with different charge distribution extending inside the material. 
The VD model, however, assumes a single excitation localized in an infinitesimally narrow region near the edge, with a propagation velocity that has to be extracted from experiments; the effect of the additional modes and of their specific charge distribution remains unspecified.

Also, mesosopic structures have a finite density of states, which is expected to renormalize the capacitive coupling between the electrodes and the Hall bar.
This is the basic idea behind the introduction of the well-known concept of quantum capacitance \cite{Buttiker, Buttiker2, Vignale}; how this additional capacitance modifies the performance of the gyrator was not quantitatively analyzed.

A deeper knowledge of the physics of the EMP is then required to gain additional insight on the response of these devices.
A lot of research has been done in the field of EMP in the QH regime.
From literature, one can distinguish two categories of EMPs depending on the smoothness of the confinement potential at the edges.

On one side, EMPs can be supported by boundaries defined by a very smooth confining potential, e.g. electrostatically defined edges. This problem is typically treated with classical hydrodynamics, neglecting corrections on the scale of the magnetic length $l_B\equiv \sqrt{e|B|/\hbar}$ \cite{Glazman, Glazman2,  JohnsonVignale}. 
Interestingly, in the same framework one can prove that edge plasmons propagate chirally without applied magnetic field in conductors with a non-zero Berry flux \cite{Rudner}, e.g. anomalous Hall materials or 2-dimensional gapped Dirac materials with light-induced valley polarization. This suggests that non-reciprocal devices could be obtained with non magnetic materials without external magnetic field; we do not analyze this case here.

On the other side, EMPs can propagate also at boundaries defined on atomic lengthscales, where corrections of the order $l_B$ are not negligible.
A classical approach for this situation was introduced in \cite{Volkov}, where the EMP problem, formulated in terms of combined Poisson and linearized continuity equations, was solved in a variety of situations with a Wiener-Hopf calculation. This approach, although very general, does not capture the physics at the QH plateus, where the transverse conductivity vanishes, i.e. $\sigma_{xx}=0$. 
A quantum generalization of the sharp edge model was proposed in \cite{Mikhailov} by linearizing the Heisenberg equation of motion of a single particle charge density operator.

Inspired by the latter research, we develop a general model  based on the linear response theory in the random phase approximation (RPA), capable to describe (by taking appropriate limits) the EMP supported by both type of edges in the QH regime.
To describe the behavior of the device, we include also an applied time-dependent voltage drive.
First, we point out the similarities and differences in the two cases for 2-dimensional electron gas (2DEG). 
Then, we find that our model  is, with few modifications, applicable also to describe EMPs in monolayer graphene, and we investigate the differences with 2DEGs. 

We then use our driven EMP model to describe a specific QH device, namely the 3-terminal gyrator introduced in \cite{Bosco}; we compare our results with the ones predicted with the VD model.

This paper is structured as follows.
In Sec. \ref{sec:model}, we introduce the EMP model. 
First, we review the eigensystem of the static Hamiltonian of independent electrons in a magnetic field, including a confinement and a mean-field (Hartree-Fock) interaction potential; we remark on the differences due to the characteristic lengthscale $w$ at which the confining potential varies.
We then focus on the effect of a time-dependent voltage drive, and by using linear response theory in RPA, we find a general equation defining the EMP charge.
At this point, we take the limits of smooth and sharp edges, and analyze the two situations.
We also modify our theory to describe EMPs in a monolayer graphene.
In Sec. \ref{sec:application}, we employ the EMP model to describe the behavior of a 3-terminal gyrator, underlining similarities and differences with VD.

\section{\label{sec:model}EMP model}

\subsection{\label{subsec:static-H0}Static Hamiltonian}

The starting point to describe the (integer) QH effect is the conventional single particle mean-field Hamiltonian \cite{QuantumHallGirvin, GirvinRev}
\begin{equation}
\hat{H}_0= \hat{H}_B +U_w(\hat{\overline{r}}) +U_i(\hat{\overline{r}}),
\label{eq:QH-hamiltonian}
\end{equation}
where $\hat{H}_B$ is the Hamiltonian of a free electron in a perpendicular magnetic field $B$, and the two scalar potentials  
$U_w$ and $U_i$ account respectively for the confinement at the edge of the material and for the mean-field (Hartree-Fock) interactions. Here, $\overline{r}=(x,y)^T$.

For 2DEGs, the magnetic field-dependent Hamiltonian $\hat{H}_B$ is \cite{Doucot}
\begin{equation}
\hat{H}_B=\hbar \omega_c \left( \hat{a}^{\dagger}\hat{a} +\frac{1}{2}\right),
\label{eq:MF-ham-2deg}
\end{equation}
where $\omega_c\equiv e B / m $ is the cyclotron frequency, and the creation and annihilation operators, $\hat{a}^{\dagger}$ and $\hat{a}$, defined by
\begin{subequations}
\label{eq:ladder-op-2DEG}
\begin{flalign}
\hat{a}^{\dagger} & \equiv \frac{l_B}{\sqrt{2}\hbar}\left(\hat{\pi}_x+ i\hat{\pi}_y \right),\\
\hat{a} & \equiv \frac{l_B}{\sqrt{2}\hbar}\left(\hat{\pi}_x- i\hat{\pi}_y \right),
\end{flalign}
\end{subequations}
satisfy the canonical commutation relation $[\hat{a},\hat{a}^{\dagger}]=1$. Here, $l_B\equiv \sqrt{\hbar/(e \lvert B \rvert)}\approx 26\mathrm{nm}/\sqrt{\lvert B \rvert/\mathrm{Tesla}}$ is the magnetic length and $\hat{\overline{\pi}}$ is the dynamical momentum
\begin{equation}
\hat{\overline{\pi}}\equiv \hat{\overline{p}}+ e \overline{A}(\hat{\overline{r}}),
\end{equation}
where $\hat{\overline{p}}$ is the crystal momentum and $\overline{A}$ is the vector potential satisfying $\overline{B}=\overline{\nabla}\times \overline{A}$.
The energy eigenvalues of the Hamiltonian in Eq. (\ref{eq:MF-ham-2deg}) are simply
\begin{equation}
\epsilon_n=\hbar \omega_c \left( n +\frac{1}{2}\right),
\label{eq:MF-ham-eigenvalue-2deg}
\end{equation}
with $n\in \mathbb{N}$ being the Landau level (LL) index.

Although our model can be straightforwardly generalized to account for an additional Zeeman splitting term, for simplicity, we neglect its effect \cite{GirvinRev}, and we consider degenerate spins.

\begin{figure}
\includegraphics[scale=0.2]{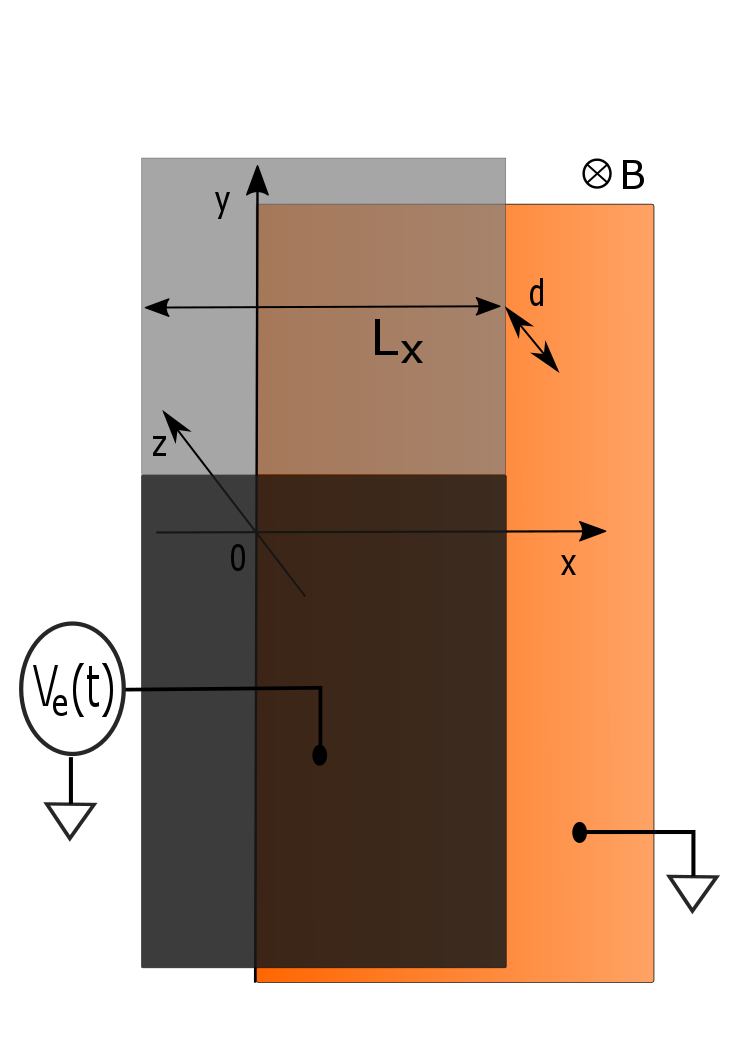}
\caption{Geometry of the EMP model. A 2-dimensional material (orange) subjected to a perpendicular magnetic field is situated in the $z=0$ plane; it extends infinitely in the $y$-direction, and it terminates at $x=0$. The material is capacitively coupled with an external electrode at $z=d$. To simplify the calculations, we will assume that the top gate is translational invariant in the $y$-direction, except that the drive changes the potential of a part of the gate only (dark grey region). For example, in the figure, we show a voltage drive of the form $V_e(t)\Theta(-y)\Theta(L_x-|x|)$.}
\label{fig:line-geometry}
\end{figure}

To proceed further and account for the confinement potential $U_w$, we need to fix the gauge of the vector potential $\overline{A}$.
In particular, since we aim to describe the EMPs propagating along a straight line, as shown in Fig. \ref{fig:line-geometry}, we choose the Landau gauge $\overline{A}(\overline{r})=\left( 0, B x, 0\right)$, which preserves the translational invariance in the $y$-direction.
Then, the eigenvalues $p_y\equiv \hbar k_y$ of the crystal momentum in the $y$-direction $\hat{p}_y$  are good quantum numbers and the eigenfunctions of the Hamiltonian in Eq. (\ref{eq:MF-ham-2deg}) are simply
\begin{equation}
\Psi(x,y)= \frac{e^{i k_y y}}{\sqrt{L_y}} \psi_n(x+k_y l_B^2),
\label{eq:MF-ham-wavefunction-2deg}
\end{equation}
with
\begin{equation}
\psi_n(x)= \frac{e^{-x^2/(2l_B^2)}}{( l_B \sqrt{\pi}2^n n! )^{1/2}}H_n(x/l_B),
\label{eq:hermite pol}
\end{equation}
and $H_n$ being the $n$th Hermite polynomials.
Here, the normalization factor $1/\sqrt{L_y}$ ($L_y$ is the length of the device in the $y$-direction) results from applying periodic boundary conditions, hence the momenta $k_y$ are quantized with steps of size $2\pi/L_y$. 

Note that the wavefunctions are centered at position $x=-k_y l_B^2$; this implies that electrons with different $y$-momenta $\hbar k_y$ are shifted in the $x$-direction. So far translations in the $x$-direction do not change the energy of the system and thus the energy eigenvalues in Eq. (\ref{eq:MF-ham-eigenvalue-2deg}) are infinitely degenerate in $k_y$.
The confinement potential $U_w$ lifts this degeneracy \cite{Hajdu}.
Let us assume that the confinement potential preserves the translational invariance in the $y$-direction, i.e.  $U_w(\overline{r})=U_w(x)$, and it has the form
\begin{equation}
U_w(x)=
\left\lbrace
\begin{array}{dd}
U_0        & x<0 \\
u(x) & 0\leq x <w \\
0          & x\geq w, \\
\end{array}
\right.
\end{equation}
with $u$ being a monotonically decreasing function interpolating continuously between the two extremes of the potential on a lengthscale $w$. 
Qualitatively, the lifting of degeneracy  in $k_y$ is easily explained: the wavefunctions centered at $-k_yl_B^2 \gg w$ experience a lower potential than the wavefunctions at $-k_yl_B^2 \approx 0$ and consequently they have lower total energy.
The detailed band structure depends on the precise form of the confining potential and typically it requires a numerical analysis. 
However, an analytical approximation for the energy eigenvalues and eigenfunctions can be found in the limits of \textit{sharp} ($w/l_B \ll 1$), and \textit{smooth} ($w/l_B \gg 1$) confinement. 

\begin{figure}
\includegraphics[scale=0.55]{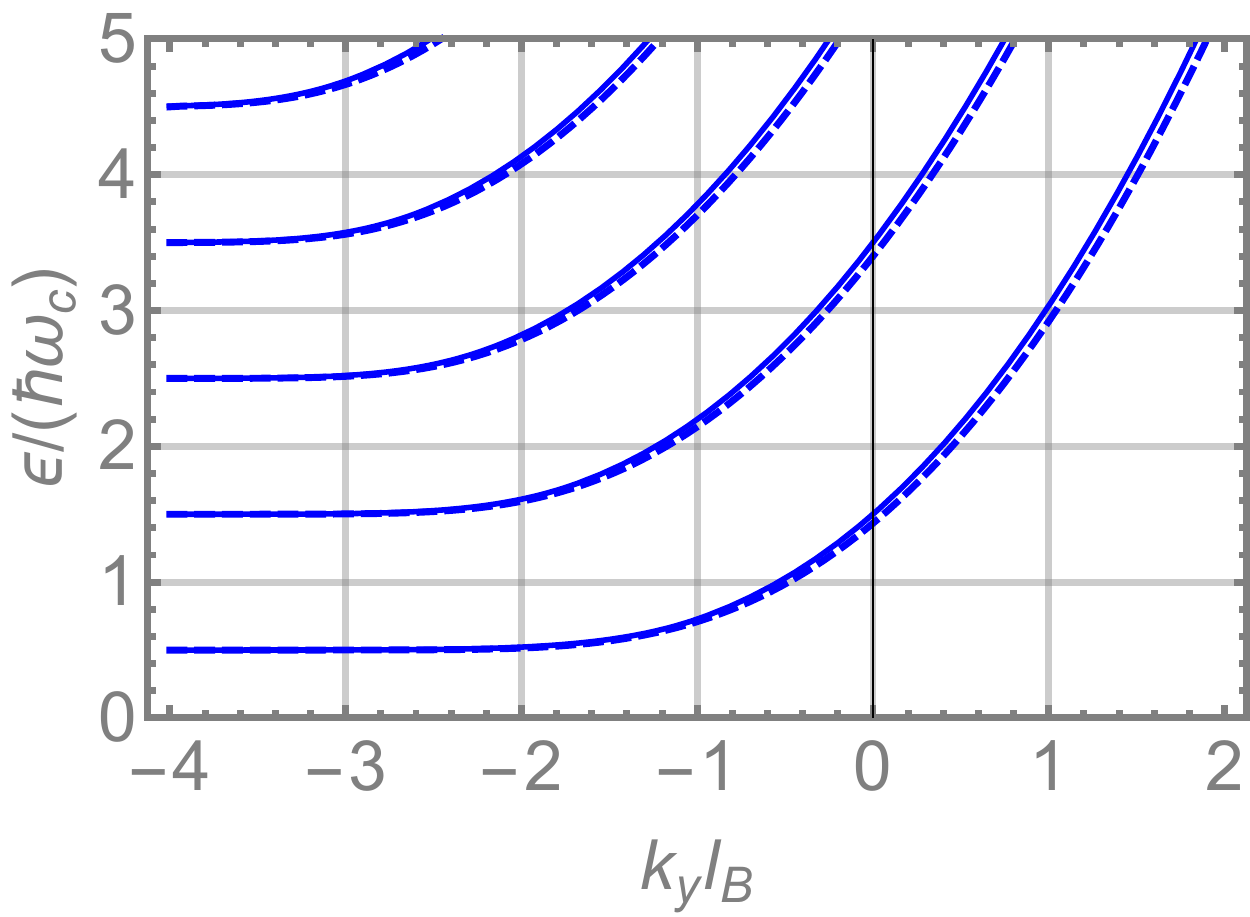}
\caption{Landau levels of a 2DEG terminated by a sharp edge potential as a function of the momentum $k_y$ (in units $1/l_B$). The energy levels are computed within the WKB approximation (solid lines) and exactly, with the dispersion relation in Eq. (\ref{eq:hermite-dispersion}) (dashed lines). Static interactions have been neglected.}
\label{fig:2deg-band-structure}
\end{figure}

If the edge potential is very sharp compared to $l_B$, one can approximate $U_w$ with a Heaviside function, $U_w(x)\approx U_0 \Theta(-x)$. Also, we consider the limit of electrons strongly confined in the material, $U_0 \gg \hbar \omega_c (\nu_0 +1/2)$, where the filling factor $\nu_0$ is the highest occupied bulk LL. 
Then, the effect of $U_w$ can be modeled by requiring the wavefunction to vanish at $x=0$.
This case has been extensively studied with different approaches \cite{Hajdu, MacDonald-Streda, Montambaux-1, Montambaux-2}. 
Although an exact analytical solution for this problem can be found in terms of the Hermite functions, as derived in Appendix \ref{app:Eigen-funct}, in this work, we use the semiclassical WKB method proposed in \cite{Montambaux-1, Montambaux-2}, that gives simple yet very good approximation for the energy eigenfunctions and eigenvalues. A comparison between the WKB and the exact band structure of a 2DEG in magnetic field is shown in Fig. \ref{fig:2deg-band-structure}.

In contrast, when the edge potential is smooth compared to $l_B$, the wavefunctions are barely perturbed by  $U_w$ and they can be approximated, to the lowest order in $l_B/w$, by the ones in Eq. (\ref{eq:MF-ham-wavefunction-2deg}), leading to the energy eigenvalues
\begin{equation}
\epsilon_n(k_y)\approx\hbar \omega_c \left( n +\frac{1}{2}\right)+U_w(k_y l_B^2).
\end{equation}

The last ingredient missing to describe the static Hamiltonian is the mean-field interaction potential $U_i$. 
Qualitatively, there are two interactions of opposite signs, that govern the edge structure in the two regimes: the long-range repulsive Coulomb (Hartree) interactions and the short-range attractive exchange (Fock) interactions \cite{Chamon}.

When the confinement potential is sharp enough, the exchange interaction dominates and the electrons simply fill the one-particle energy bands in Fig. \ref{fig:2deg-band-structure} up to the Fermi energy $\epsilon_F$.
In this case, there are exactly $\nu_0$ states at the Fermi energy; each of these states has a well-defined Fermi momentum $k_F^n$ ($n$ labels the LLs), and corresponds to a current-carrying channel in Landauer language.
How this picture is modified for fractional filling factors is described in \cite{MacDonald, Brey}. 

In contrast, it is well-known that for very smooth confinement potentials, e.g. electrostatic confinement, the long-range Coulomb repulsive force dominates and the edge undergoes reconstruction. The structure of the edge in this case has been extensively studied focusing on the electronic density and neglecting all the details at lengthscales of the order of the magnetic length. 
From a semiclassical electrostatic approach \cite{Beenakker, CSG}, the electron density at the edge shows an alternating pattern of compressible and incompressible strips.
In particular, Coulomb interactions flatten the energy bands and instead of having single current-carrying states with a unique Fermi momentum $k_F^n$, at the Fermi energy there is a set of quasi-degenerate energy eigenstates for each LL. These sets correspond to the compressible current-carrying strips and they are spatially separated by narrow incompressible insulating strips that appear at integer local filling factor $\nu(x)$.
This picture was confirmed to hold also at fractional filling factors both by DFT calculations, including exchange interactions at low temperature \cite{Ferconi}, and by composite fermions approach \cite{Chklovskii}. 

The transition between the two limits has been studied in detail for the integer QH effect within an Hartree-Fock mean-field theory \cite{Chamon} and for the fractional QH case with a composite fermions approach \cite{Chklovskii}. The cross-over between the two limits is estimated to occur when $w$ is of the order of the magnetic length $l_B$.

\subsection{\label{subsec:driven-case} Driven Hamiltonian}

Imagine now to perturb the system (a 2DEG in the plane $z=0$) described by the Hamiltonian $\hat{H}_0$ by a time-dependent voltage drive $V_e(t)$, applied to an electrode in a parallel plane $z=d$, as shown in Fig. \ref{fig:line-geometry}.
The drive is assumed to be slow enough for the retardation effects to be negligible, a condition typically met in the microwave domain. For simplicity, we neglect also fringing fields: the voltage applied in position $z=d$ preserves its spatial distribution at $z=0$. The effect of fringing fields is briefly discussed in Appendix \ref{app:Fringing}.
The drive perturbs the electron density from its equilibrium value $\rho_0$, causing a time-dependent rearrangement of charges; the change in density adds a significant Coulomb energy cost  $U_{\rho}$, which should be included in the Hamiltonian.

In a time-dependent Hartree-Fock approximation, this additional energy term is included in the total \textit{screened} scalar potential $U$ \cite{Vignale}
\begin{equation}
U(\overline{r},t)\equiv - e V_e(\overline{r},t)+ U_{\rho}(\overline{r},t),
\label{eq:screened-potential}
\end{equation}
with $\overline{r}=(x,y)^T$.
As  $U_{\rho}$ re-adapts self-consistently to the perturbed charge density, this additional term leads to a complicated set of nested non-linear integral equations.

The dynamics of the electron density can be simplified in the framework of linear response theory by making use of the random phase approximation (RPA).
First, we neglect exchange interactions, and we model $U_{\rho}$ by inverting the electrostatic Poisson equation
\begin{equation}
U_{\rho}(\overline{r},t)=e\int_{\mathbb{R}^2} d \overline{r}' G_0(\overline{r},\overline{r}')\left(\rho(\overline{r}',t)-\rho_0(\overline{r}')\right).
\label{eq:inverted-Poisson-equation}
\end{equation}
Here, $\rho$ is the non-equilibrium expectation value of the charge density operator, and $G_0$ is the electrostatic Green's function of the 3-dimensional Poisson operator in Eq. (\ref{eq:Greens-top-gate-3d}), evaluated at $z=0$. 

We work in the frequency domain $t\rightarrow\omega$, and we assume that the external voltage is small enough for the induced charge density to be linear in $V_e$. 
This assumption allows one to study the first-order charge density perturbation,
\begin{equation}
\rho_1(\overline{r},\omega) \equiv \rho(\overline{r},\omega)-\rho_0(\overline{r}),
\end{equation}
in terms of linear response functions, depending only on equilibrium averages over the eigenfunctions of $\hat{H}_0$.

In particular, we introduce the \textit{proper} density-density response function $\tilde{\chi}_{\rho \rho}$ \cite{Vignale}, defined by
\begin{equation}
\rho_1(\overline{r},\omega)\equiv e \int_{\mathbb{R}^2}d\overline{r}' \tilde{\chi}_{\rho \rho} (\overline{r},\overline{r}', \omega) U(\overline{r}',\omega).
\label{eq:linear-charge-rpa}
\end{equation}
In RPA, $\tilde{\chi}_{\rho \rho}$ is given by, in Lehmann representation, \cite{Vignale, Rammer}
\begin{multline}
\tilde{\chi}_{\rho \rho}(\overline{r},\overline{r}',\omega)=\\
\sum_{\alpha,\beta} \frac{f_F (\epsilon_\beta)-f_F (\epsilon_\alpha)}{\epsilon_\beta-\epsilon_\alpha +\hbar (\omega + i \eta) }
 \Psi_{\alpha}(\overline{r})\Psi^*_{\beta}(\overline{r})\Psi_{\beta}(\overline{r}')\Psi^*_{\alpha}(\overline{r}').
\label{eq:RPA-response-function}
\end{multline}
Here, $f_F$ is the Fermi distribution and the imaginary part of the frequency $\eta\ll \omega$ can be interpreted as a phenomenological decay rate due to the coupling to the environment. 
The indexes $\alpha,\beta$ collect all the quantum numbers associated to $\hat{H}_0$, in this case the LL number $n$ and the crystal momentum $k_y$; $\Psi_{\alpha}$ and $\Psi_{\beta}$ are the eigenfunctions of $\hat{H}_0$ with eigenvalues $\epsilon_{\alpha}$ and $\epsilon_{\beta}$ respectively.

Introducing the matrix decomposition for the linearized charge density 
\begin{equation}
\rho_1(\overline{r},\omega)=\sum_{\alpha, \beta} \rho_{\alpha \beta} (\omega) \Psi_{\alpha}(\overline{r})\Psi^*_{\beta}(\overline{r}),
\label{eq:matrix-decomp-rho}
\end{equation}
and substituting Eq. (\ref{eq:RPA-response-function}) into Eq. (\ref{eq:linear-charge-rpa}), one gets
\begin{equation}
\rho_{\alpha \beta}(\omega)=e\frac{f_F (\epsilon_\beta)-f_F (\epsilon_\alpha)}{\epsilon_\beta-\epsilon_\alpha +\hbar (\omega + i \eta) } U_{\alpha \beta}(\omega),
\label{eq:matrix-element-rho}
\end{equation}
where we have introduced the screened potential matrix element
\begin{equation}
U_{\alpha \beta}(\omega)=\int_{\mathbb{R}^2} d  \overline{r}'\Psi^*_{\alpha}(\overline{r}')\Psi_{\beta}(\overline{r}') U(\overline{r}',\omega).
\label{eq:us-matrix-full}
\end{equation}

In the undriven case, i.e. $V_e=0$, $U=U_{\rho}$, Eq. (\ref{eq:matrix-element-rho}) was obtained in \cite{Mikhailov} by linearizing the Heisenberg equation of motion of $\rho$ for small density perturbation.

Now, we use the translational invariance in $y$-direction of $\hat{H}_0$, which allows the factorization
\begin{equation}
\Psi_{\alpha}(\overline{r})= \frac{e^{i k_y y}}{\sqrt{L_y}} \psi_{\alpha}(x).
\label{eq:Psi-factorization}
\end{equation}

A few remarks are in order here. First, $\Psi_{\alpha}$ is the eigenfunction of $\hat{H}_0$ and it is different from $\Psi$ in Eq. (\ref{eq:MF-ham-wavefunction-2deg}), which is the eigenfunction of the free-electron Hamiltonian in Eq. (\ref{eq:MF-ham-2deg}).
In addition, the driving voltage $V_e$ is a function of $y$, so that, although the static density is still assumed to be constant in the $y$-direction, the perturbation $\rho_1$ is not.

To proceed further, we Fourier transform the $y$-coordinate, $y \rightarrow q_y$, and, following \cite{Mikhailov}, we use some physically reasonable approximations.
First, we assume that the temperature is low enough to have fully developed QH plateaus, i.e. $k_B T \ll \hbar \omega_c$, and we consider that the size of the sample is much greater than all the other lengthscales, which allows us to take the thermodynamic limit, $L_y\rightarrow\infty$. In this limit, the momentum quantum number $k_y$ becomes a continuous parameter and, in our notation, we promote it to be an argument of the functions.
Moreover, we focus only on low-energy excitations, with $\omega\ll\omega_c$, whose variations in the $y$-direction are much smoother than in the $x$-direction, i.e. $\mathrm{max}(w, l_B) q_y \ll 1$.

With these assumptions, and decoupling the $y$ and $x$ directions by introducing the quantities  $p_n\left(k_y, q_y, \omega\right)$ and $V_{n}\left(k_y,q_y,\omega \right) $, respectively defined by
\begin{equation}
\rho_1\equiv \sum_n \int_{\mathbb{R}} dk_y \frac{\partial f_F\left(\epsilon_n(k_y)\right)}{\partial k_y} p_n\left(k_y, q_y, \omega\right) \left| \psi_n(x, k_y)\right|^2,
\label{eq:pn-def}
\end{equation}
and
\begin{equation}
V_{n}\left(k_y,q_y,\omega \right) \equiv \int_{\mathbb{R}} dx V_e(x,q_y,\omega) \left| \psi_n(x, k_y)\right|^2,
\label{eq:Ve-matrix}
\end{equation}
we get the self-consistent equation
\begin{multline}
-\left(\omega+ i \eta\right) p_{n}\left(k_y, q_y,\omega \right)=\frac{q_y e^2}{2 \pi \hbar } V_{n}\left(k_y,q_y,\omega \right) +\\
q_y\sum _m\int _\mathbb{R} dk'_y
\mathcal{M}_{n m}\left( k_y, k'_y, q_y\right) p_{m}\left(k'_{y}, q_y,\omega \right).
\label{eq:motion-eq-general}
\end{multline}
Here, the quantity
\begin{equation}
\begin{split}
\mathcal{M}_{n m}\left( k_y, k'_y, q_y\right) \equiv & v^q_{n}(k'_y)\delta(k_y-k'_y)\delta_{n m}-\\
 & v^c_{n m}\left( k_y, k'_y, q_y\right) \frac{\partial  f_F \left(\epsilon_{m}(k'_y)\right)}{\partial k'_y},
 \end{split}
 \label{eq:m-sum-velocity}
\end{equation}
includes the two velocities that contribute to the motion of the excitation:
first, the group velocity of a wavepacket centered at momentum $k_y$ (quantum contribution),
\begin{equation}
v^q_n(k_y) \equiv \frac{\partial \epsilon_n(k_y)}{\hbar\partial k_y},
\label{eq:quantum-velocity}
\end{equation}
and second, the electrostatic contribution due to the self-consistent rearrangement of charges (classical contribution)
\begin{multline}
v^c_{n m}\left( k_y, k'_y, q_y\right) \equiv  \frac{e^2}{\hbar}\int_{\mathbb{R}^2} dx dx' G_0(x,x',q_y) \times \\
 \left| \psi_n(x, k_y)\right|^2 \left| \psi_m(x', k'_y)\right|^2 .
 \label{eq:classical-velocity}
\end{multline}

The derivations leading to Eq. (\ref{eq:motion-eq-general}) are reported in Appendix \ref{app:motion-eq}.

Note that the sum of two velocity contributions is expected from the well-known concept of quantum capacitance \cite{Buttiker, Buttiker2, Vignale}.
In fact, if we consider a simple circuit model describing our system, as in \cite{Viola-DiVincenzo, Glattli}, the total velocity of the collective edge-excitations can be related to the inverse of an effective electrochemical capacitance. In mesoscopic devices, this quantity is modeled by a geometrical capacitance dependent on the physical distance of the electrode $d$, in series with a quantum capacitance accounting for the density of states, i.e. $\propto (\partial\epsilon/\partial k_y)^{-1}$; this agrees with our Eq. (\ref{eq:m-sum-velocity}).
The important connection between plasmon velocities and capacitances  will be developed more in Sec. \ref{subsec:VD-RPA}.

We will now examine in detail Eq. (\ref{eq:motion-eq-general}) in the smooth  and in the sharp edge limit.

\subsection{\label{subsec:smooth-edges}Smooth edges}

We now employ Eq. (\ref{eq:motion-eq-general}) in the limit of smooth edges, $w /l_B \gg 1 $.
In this case, $\psi_n(x)$ is proportional to a Gaussian function centered at position $x=-k_y l_B^2$, with standard deviation approximately $l_B\sqrt{n}$, see Eq. (\ref{eq:hermite pol}) and relative discussion; neglecting all the details at lengthscales of the order of the magnetic length, one can thus approximate the absolute value squared of the wavefunctions with shifted Dirac delta functions.
In addition, as discussed in Sec. \ref{subsec:static-H0}, Coulomb interactions flatten the energy bands $\epsilon_n(k_y)$ at the Fermi energy, thus we discard the quantum velocities defined by Eq. (\ref{eq:quantum-velocity}).

With these two assumptions, one obtains the integral equation
\begin{multline}
(\omega + i\eta) \rho_1(x,q_y,\omega)=\frac{\partial \rho_0(x)}{\partial x}
 \left(- \frac{e^2 q_y}{m \omega_c} V_e(x,q_y, \omega) + \right.\\
\left. \frac{2\pi e^2 q_y}{m \omega_c} \int_{\mathbb{R}}dx'\rho_1(x',q_y,\omega)G_0(x,x',q_y)\right),
\label{eq:motion-smooth-edges}
\end{multline}
as derived in Appendix \ref{app:smooth-edge}.

Let us first compare this result with literature. 
The problem of low-energy, smooth excitations supported by smooth edges of a QH liquid has often been studied within an hydrodynamic approach.
Aleiner and Glazman (AG) \cite{Glazman, Glazman2} combine the Euler equation for compressible electron liquid, including a potential term of the same form of $U_{\rho}$ in Eq. (\ref{eq:inverted-Poisson-equation}) with the linearized continuity equation, and they find a self-consistent integro-differential equation for the charge density.

In the high magnetic field limit $\omega/\omega_c \ll 1$ and in the undriven case, i.e. $V_e=0$, their Fredholm integral equation coincides with our Eq. (\ref{eq:motion-smooth-edges}) (up to a minus sign due to different conventions for the Fourier transform in time), when we use the free-space electrostatic Green's function
\begin{equation}
G_0^f(x,x',q_y)=\frac{K_0(\left|q_y\right|\left|x-x'\right|)}{4 \pi^2 \epsilon_S},
\label{eq:free-e-GF}
\end{equation}
where $K_0$ is the modified Bessel function and $\epsilon_S$ is the dielectric constant of the medium.
Note that this equivalence is expected as the RPA preserves the continuity equation \cite{KurasawaRPA}.
Eq. (\ref{eq:free-e-GF}) can be derived by Fourier transforming the $y$-coordinate in Eq. (\ref{eq:Greens-top-gate-3d}) (evaluated at $z=0$) and by taking the limit $d\rightarrow\infty$.

AG found that smooth edges support infinitely many branches of plasmonic excitations, each with a different number of nodes in the $x$-direction. The number of nodes is strictly correlated with the velocity of the plasmons: excitations with fewer nodes, are faster. In particular, the fastest mode has a logarithmic dispersion relation $\omega(q_y)$, with velocity diverging for $q_y\rightarrow0$.
This logarithmic behavior is well-known, both theoretically \citep{Volkov, Mikhailov} and experimentally \cite{Ashoori, Balaban}, and it is related to the logarithmic divergence of the free-space electrostatic Green's function in Eq. (\ref{eq:free-e-GF}) for small $q_y$.

In our geometry, however, the top gate at a distance $d \ll 1/q_y$, slows the plasmons, and in particular the presence of a positive image charge at position $z=2d$ straightforwardly modifies the Green's function in Eq. (\ref{eq:free-e-GF}) to
\begin{multline}
G_0(x,x',q_y)=\frac{1}{4 \pi^2 \epsilon_S}\left[K_0(\left|q_y\right|\left|x-x'\right|) - \vphantom{ K_0\left(\left|q_y\right|\sqrt{(x-x')^2+4d^2}\right)} \right. \\
\left. K_0\left(\left|q_y\right|\sqrt{(x-x')^2+4d^2}\right)\right].
\label{eq:ic-Gf}
\end{multline}

Expanding Eq. (\ref{eq:ic-Gf}) for small $q_y$, consistent with the smooth excitation approximation, the logarithmic divergences of the two Bessel functions cancel out and we obtain the finite limit 
\begin{equation}
\lim_{q_y\rightarrow 0}G_0=\frac{1}{8 \pi^2 \epsilon_S}\log\left(1+\left(\frac{2d}{x-x'}\right)^2 \right),
\label{eq:linear-top-gate-Greens-function}
\end{equation}
which presents the typical logarithmic behavior expected from 2-dimensional electrostatics \cite{Macdonald-Rice}.

When no voltage is applied, a solution of Eq. (\ref{eq:motion-smooth-edges}) with the kernel given in Eq. (\ref{eq:linear-top-gate-Greens-function}) was found by Johnson and Vignale (JD) \citep{JohnsonVignale}.
JD used a linear static charge density $\rho_0$, including both compressible and incompressible strips, and they solved the problem within a local capacitance approximation (LCA), valid for $d/w\ll 1$. 
Among other things, they proved that the presence of incompressible strips gives a negligible contribution to the EMP velocities, and thus we will neglect them here.

Although the LCA greatly simplifies the problem, with this approach much information on the multipole modes is lost. Instead, we deal with the problem following the strategy of AG, and we use an orthogonal polynomial decomposition to decouple the $x$ and $y$-directions.
We introduce their static charge density \cite{Glazman}
\begin{equation}
\rho_0(x)=\Theta(x)\frac{2 n_0 }{\pi}\tan^{-1}\sqrt{\frac{x}{w}},
\label{eq:static-smooth-charge}
\end{equation}
with $n_0$ being the bulk density of electrons, and we factorize the excess charge density as
\begin{equation}
\rho_1(x, q_y, \omega)=\sum_j c_j(q_y,\omega)R_{j}(x),
\label{eq:factorization-smooth1}
\end{equation}
where we have defined 
\begin{subequations}
\begin{flalign}
R_{0}(x)&\equiv \frac{\Theta(x)}{\pi \sqrt{x/w}(x+w)}, \\
R_{j}(x)&\equiv \frac{\sqrt{2} \Theta(x)}{\pi \sqrt{x/w}(x+w)}   T_{2j}\left(\frac{1}{\sqrt{1+x/w}}\right),
\end{flalign}
\end{subequations}
with $T_j$ being the $j$th Chebyshev polynomial of the first kind \cite{Stegun}.
Combining now Eqs. (\ref{eq:motion-smooth-edges}), (\ref{eq:linear-top-gate-Greens-function}) and (\ref{eq:factorization-smooth1}), and using the orthogonality of the Chebyshev polynomials, one gets the equation for the vector of coefficients $\overline{c}(q_y,\omega)\equiv[c_0(q_y,\omega),c_1(q_y,\omega),...]^T$
\begin{equation}
(\omega+i \eta)\overline{c}(q_y,\omega)= q_y \hat{\mu}\overline{c}(q_y,\omega)+q_y \overline{V}(q_y,\omega).
\label{eq:motion-smooth-c-fourier}
\end{equation}
Here, $\hat{\mu}$ is a symmetric real matrix with units of velocity and with elements
\begin{multline}
\mu_{i j}\equiv \gamma_{i j}\frac{2 n_0 e^2}{\pi^3 m \omega_c \epsilon_S} \int_0^1 ds \frac{ T_{2i}(s)}{\sqrt{1-s^2}}\int_0^1ds'\frac{T_{2j}(s')}{\sqrt{1-s'^2}} \times \\
\log \left( 1+\left(\frac{2d}{w}\frac{s^2 s'^2}{s^2 -s'^2}\right)^2 \right),
\label{eq:electrostatic-mu}
\end{multline}
with 
\begin{equation}
\gamma_{ij}\equiv\left\lbrace
\begin{array}{cc}
1/2 & i=j=0,\\
1/\sqrt{2} & i=0 \veebar j=0, \\
1 & \mathrm{otherwise},
\end{array}
\right.
\end{equation}
and $\overline{V}$ is a vector with elements 
\begin{subequations}
\label{eq:voltage-vectorsmooth}
\begin{flalign}
V_{0} &\equiv -\frac{2 n_0 e^2}{\pi m \omega_c}\int_0^1 \frac{ ds}{\sqrt{1-s^2}} V_e\left(\frac{w}{s^2}-w, q_y, \omega\right), \\
V_{j} &\equiv -\frac{2\sqrt{2} n_0 e^2}{\pi m \omega_c}\int_0^1 ds \frac{ T_{2j}(s)}{\sqrt{1-s^2}} V_e\left(\frac{w}{s^2}-w, q_y, \omega\right).
\end{flalign}
\end{subequations}

Note that Eq. (\ref{eq:motion-smooth-c-fourier}) corresponds to the linear system of coupled partial differential equations
\begin{equation}
\frac{\partial\overline{c}(y, t)}{\partial t}+\eta \overline{c}(y, t)= \hat{\mu} \frac{\partial\overline{c}(y, t)}{\partial y}+\frac{\partial \overline{V}(y, t)}{\partial y},
\label{eq:motion-smooth-c}
\end{equation}
that can always be decoupled by the unitary transformation $\hat{M}$ that diagonalizes $\hat{\mu}$. Conventionally, $\hat{M}$ is chosen to be the matrix containing the column eigenvectors, properly normalized to satisfy $\hat{\mu}=\hat{M}\hat{v}\hat{M}^T$, with $\hat{v}$ being the matrix of eigenvalues.
 
We are now able to define the $j$th plasmon velocity $v_j$ and its wavefunction in the $y$-direction $u_j(y,t)$, as the $j$th eigenvalue of $\hat{\mu}$ and the linear combination of coefficients $c_i$ given by
\begin{equation}
u_j(y,t)\equiv \sum_i M_{ij} c_i(y,t).
\label{eq:plasmon-y-smooth}
\end{equation}
Hence, we reduce the system in Eq. (\ref{eq:motion-smooth-c}) to
\begin{equation}
\frac{\partial u_j(y, t)}{\partial t}+\eta u_j(y, t)= v_j \frac{\partial u_j(y, t)}{\partial y}+\hat{M}^T\frac{\partial \overline{V}(y, t)}{\partial y}.
\label{eq:motion-smooth-eigen}
\end{equation}

Combining Eqs. (\ref{eq:factorization-smooth1}) and (\ref{eq:plasmon-y-smooth}), the linearized charge density can be written as a sum of independent plasmonic contributions
\begin{equation}
\rho_1(x,y,t)=\sum_j g_j(x)u_j(y,t),
\label{eq:plasmon-full-sum}
\end{equation}
with $u_j(y,t)$ satisfying the equation of motion (\ref{eq:motion-smooth-eigen}) and 
\begin{equation}
 g_j(x)\equiv\sum_i M_{ij}R_i(x).
\end{equation}

It is now clear that both the static and dynamic components of the plasmons $g(x)$ and $u(y,t)$ are strictly related to the eigenvalues and eigenvectors of the velocity matrix $\hat{\mu}$.
In the following, we sort the eigenvalues $v_j$ from the highest to the lowest.

Note that the plasmon velocities $v_j$ have a natural scale $v_p\equiv \sigma_{xy}/(2\pi \epsilon_S)$, where $\sigma_{xy}=e n_0/B$ is the high-magnetic field conductivity in the Drude model \citep{QuantumHallGirvin}.
This velocity scale also agrees with the well-known classical Wiener-Hopf calculation of the dispersion of the EMPs in \cite{Volkov}.

\begin{figure}
\includegraphics[scale=0.7]{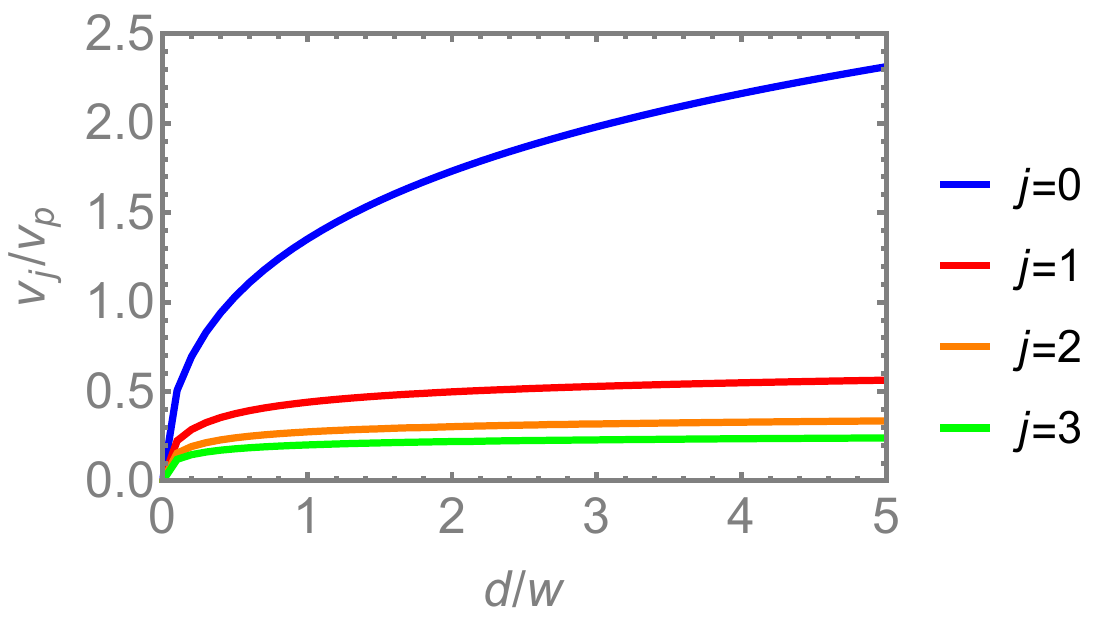}
\caption{\label{fig:velocity-smooth} Velocity of EMPs supported by smooth edges in 2DEG as a function of the distance $d$ of the top gate. The velocities are expressed in units of $v_p\equiv\sigma_{xy}/(2\pi \epsilon_S)$, while $d$ is in units of the smoothness parameter $w$, defined in Eq. (\ref{eq:static-smooth-charge}). In this calculation, we are neglecting the effects of the incompressible strips and of backscattering.}
\end{figure}

Fig. \ref{fig:velocity-smooth} shows how the plasmon velocities, normalized over $v_p$, change as a function of the distance $d$ between the top-gate and the 2DEG normalized over the $w$, defined in Eq. (\ref{eq:static-smooth-charge}).
When $d/w$ increases, the plasmons become faster and, in particular, the velocity of the fastest mode diverges in the free-space limit, $d\rightarrow\infty$, while the velocities of the others saturate to a finite value, consistent with \cite{Glazman}.

In the approximations used, although the detailed structure of $g_j(x)$ depends on the position of the top gate $d$ and on the lengthscale $w$, the $j$th plasmon mode has always $j$ nodes, as expected.

We are now able to examine the motion in the $y$-direction, defined by Eq. (\ref{eq:motion-smooth-eigen}), which is a linear partial differential equation with a damping term and an external drive.
For simplicity, we assume that the electrode driving the excitation extends in the $x$-direction for a lengthscale much greater than $w$, such that we can consider the applied voltage constant in $x$, $V_e(x,y,t)\approx V_e(y,t)$. With this approximation, because of the orthogonality of the Chebyshev polynomials, only the $0$th component of the voltage vector in Eq. (\ref{eq:voltage-vectorsmooth}) is nonzero, and Eq. (\ref{eq:motion-smooth-eigen}) becomes
\begin{equation}
\frac{\partial u_j(y, t)}{\partial t}+\eta u_j(y, t)= v_j \frac{\partial u_j(y, t)}{\partial y}+a_j \frac{ \partial V_e(y, t)}{\partial y}.
\label{eq:motion-smooth-eigen-aj}
\end{equation}
Here, $a_j$ is a parameter defined by
\begin{equation}
a_j\equiv-\frac{e^2 n_0}{m \omega_c}M_{0j}=-\sigma_{xy}M_{0j},
\end{equation}
that quantifies the coupling of the $j$th plasmon to the applied potential. 

Let us now focus on the simple yet meaningful case of an external potential of the form
\begin{equation}
V_e(y,t)=V\Theta(-y)\Theta(t)e^{i \omega_0 t}.
\label{eq:driving-harm}
\end{equation} 

Assuming equilibrium at $t=0$, one obtains from Eq. (\ref{eq:motion-smooth-eigen-aj})
\begin{equation}
u_j(y, t)=V\frac{a_j}{v_j}e^{\eta y/v_j}e^{i \omega_0 (t+y/v_j)}\left[\Theta(y)-\Theta(y+v_j t) \right]\Theta(t).
\end{equation}

\begin{figure}
\includegraphics[scale=0.6]{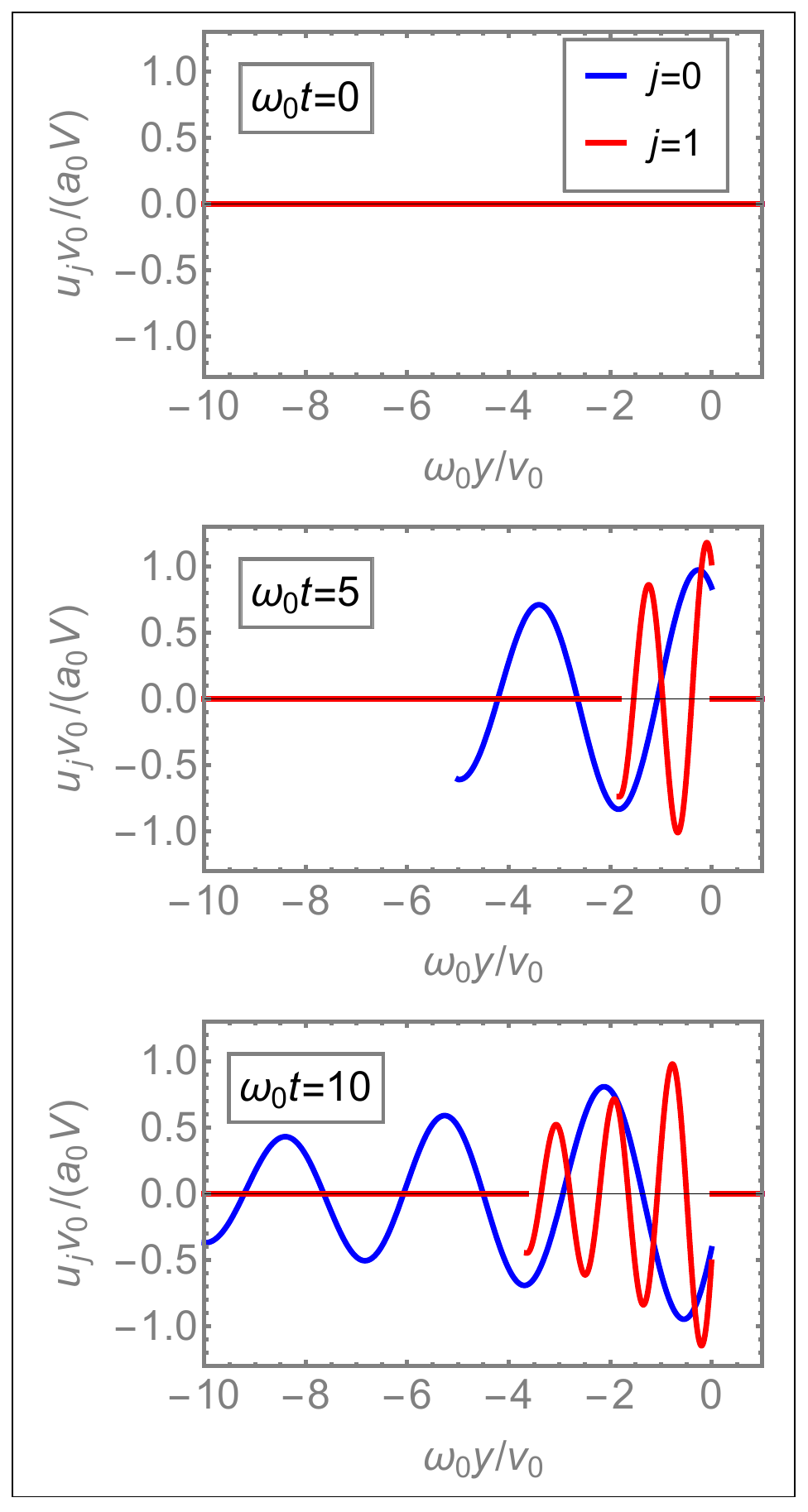}
\caption{\label{fig:chargey-smooth} Motion of the $u_j(y,t)$ component of the EMP charge under the effect of the drive $V_e$ defined as the real part of Eq. (\ref{eq:driving-harm}). All the quantities are conveniently normalized to be dimensionless. In the plot, we are showing only the first two EMP modes. We used $d/w=0.5$ and $\eta/\omega_0=0.1$. }
\end{figure}

At $t\geq 0$ plasmon waves are launched in the negative $y$-direction with different amplitudes $a_j/v_j$, velocities $v_j$ and decay length $v_j/\eta$, as shown in Fig. \ref{fig:chargey-smooth}.

A few remarks are in order here.
Although we are including a phenomenological damping rate $\eta$, we are neglecting the change in EMP velocity and distribution in the $x$-direction due to scattering.
This approximation is justified, from standard QH theory \cite{Hajdu, QuantumHallGirvin, Vignale}, when the Fermi energy is  well between two bulk LLs. In this case, the conducting states are localized at the edges of the material and back scattering is suppressed (this holds even in the presence of impurities in the material \cite{Laughlin}). 
How a magnetic field dependent scattering timescale $\tau$ modifies the EMP velocity in the smooth edge case, was examined with a semiclassical (hydrodynamic) model by JD \cite{JohnsonVignale}. When the Fermi energy meets a bulk-LL, backscattering reduces the EMP velocity, leading to downward cusps at the corresponding magnetic fields. This might explain the frequency behavior of the device in \cite{Reilly}. The latter effects seem, however, to be appreciable for rather low scattering time, $(\omega_c \tau)_{\mathrm{min}}\approx 1$, and we will neglect them here.


\subsection{\label{subsec:sharp-edges}Sharp edges, 2DEG}

We now examine the dynamics of the edge-magneto plasmons supported by sharp edges of a 2DEG.
As discussed in Sec. \ref{subsec:static-H0}, for confining potential varying with a lengthscale $w\lesssim l_B$, and assuming small thermal energy compared to the cyclotron energy, $k_BT\ll \hbar\omega_c$, the electrons fill all the states up to a Fermi momentum $k_F^n$ unique for each of the LLs.

In this case, Eqs. (\ref{eq:pn-def}) and (\ref{eq:motion-eq-general}) reduce respectively to
\begin{equation}
\rho_1(x,q_y,\omega)=\sum_{n=0}^{\nu_0-1}p_n(q_y,\omega)\left| \psi_n(x)\right|^2,
\label{eq:charge-sharp1}
\end{equation}
and
\begin{multline}
\left(\omega+ i \eta\right) p_{n}\left( q_y,\omega \right)=-q_y\frac{e^2}{2 \pi \hbar } V_{n}\left(q_y,\omega \right) +\\
q_y\sum _{m=0}^{\nu_0-1}
\mu_{n m}\left(q_y\right) p_{m}\left(q_y,\omega \right),
\label{eq:motion-sharp}
\end{multline}
where, to simplify the notation, we made the Fermi momentum dependence implicit, e.g. $p_{n}\left( q_y,\omega \right)\equiv p_{n}\left( k_F^n, q_y,\omega \right)$. The matrix element $\mu_{n m}$ has units velocity and it is defined by
\begin{equation}
\mu_{n m}\left(q_y\right) \equiv  v^q_{n}\delta_{n m}+ v^c_{n m}\left( q_y\right).
\label{eq:velocity-sharp}
\end{equation}

This velocity matrix is equivalent to the one obtained in \cite{Mikhailov} in the undriven case.
It is worth remarking here that the structure of the velocity matrix  $\hat{\mu}$, involving a sum of an electrostatic and a quantum contribution, is consistent with the concept of quantum capacitance, discussed in Sec. \ref{subsec:driven-case}.


For a top gate geometry and in the same smooth variations in $y$-coordinate approximation discussed in Sec. \ref{subsec:smooth-edges}, the electrostatic velocity matrix elements are independent of $q_y$ and given by
\begin{multline}
v^c_{n m}=v_p\int_{\mathbb{R}^2}dxdx'\left| \psi_n(x)\right|^2\left| \psi_m(x')\right|^2\times \\ 
\log\left(1+\left(\frac{2 d}{x-x'}\right)^2\right),
\end{multline}
where the characteristic velocity scale $v_p\equiv c\alpha/(2\pi \epsilon^*_S)$ depends on the speed of light in vacuum $c$, the fine structure constant $\alpha$ and the dimensionless medium permittivity $\epsilon^*_S$. Note that $v_p$, if expressed in terms of the quantum Hall conductivity $\sigma_{xy}\equiv e^2\nu_0/h$ \cite{QuantumHallGirvin}, is very similar to the characteristic velocity for smooth edges, in fact, $v_ p\propto\sigma_{xy}/(\nu_0\epsilon_S)$.

Comparing Eqs. (\ref{eq:motion-smooth-c}) and (\ref{eq:motion-sharp}), one can easily verify that the motion of the plasmons in $y$-direction is governed in both sharp and smooth edge case by a very similar system of linear partial differential equations.
Proceeding as before, we introduce the unitary transformation $\hat{M}$ that diagonalizes the symmetric velocity matrix $\hat{\mu}$ and we assume that the external potential is constant in $x$.
The total excess charge in Eq. (\ref{eq:charge-sharp1}), can then be rewritten in terms of plasmons 
\begin{equation}
\rho_1(x,y,t)=\sum_{j=0}^{\nu_0-1} g_j(x)u_j(y,t),
\end{equation}
where, here, $g_j$ is defined by
\begin{equation}
g_j(x)\equiv \sum_{i=0}^{\nu_0-1}M_{ i j} \left| \psi_i(x)\right|^2,
\label{eq:chargex-sharp}
\end{equation}
and $u_j(y,t)$ obeys the equation of motion (\ref{eq:motion-smooth-eigen-aj}), where $v_j$ is identified as the $j$th eigenvalue of the velocity matrix in Eq. (\ref{eq:velocity-sharp}) and
\begin{equation}
a_j\equiv-\frac{e^2}{2\pi \hbar}\sum_{i=0}^{\nu_0-1}M_{i j}=-\sigma_{xy}\frac{1}{\nu_0}\sum_{i=0}^{\nu_0-1}M_{i j}.
\label{eq:coupling-sharp}
\end{equation}

It is now worth remarking on some differences between smooth and sharp edges.
First, sharp edges only support a finite number $\nu_0$ of excitations, while smooth edges have an infinite spectrum of modes.
Moreover, although the spatial distribution in the $x$-direction changes significantly in the two cases, the propagation in the $y$-direction is defined by a system of partial differential equations of the same structure. 
The velocities $v_j$ and coupling $a_j$ are different for sharp and smooth edges, although the characteristic scale of $a_j$ and of the electrostatic velocities can be expressed in a similar way in terms of the high-magnetic field Hall conductivity $\sigma_{xy}$, in its classical and quantized form, respectively.

To estimate the EMP velocities, we find the band structure and the eigenfunctions of the static Hamiltonian in the  WKB approximation described in \citep{Montambaux-1, Montambaux-2}.
We assume that the edge terminates abruptly at $x=0$, and neglect the mean-field interaction potential $U_i$, and so the eigensystem is obtained by imposing Dirichlet boundary conditions to Eq. (\ref{eq:MF-ham-2deg}).
We also account for an additional factor $2$ in the EMP velocity due to the spin-degeneracy. In fact, including the degenerate spin degree of freedom, each element of the velocity matrix $\mu_{nm}$ transforms into a 2x2 matrix, with elements all equal to $\mu_{nm}$. Then, half of the eigenvalues of the new velocity matrix is zero and the other half is twice the eigenvalues computed without considering spin.

\begin{figure}
(a)\includegraphics[scale=0.48]{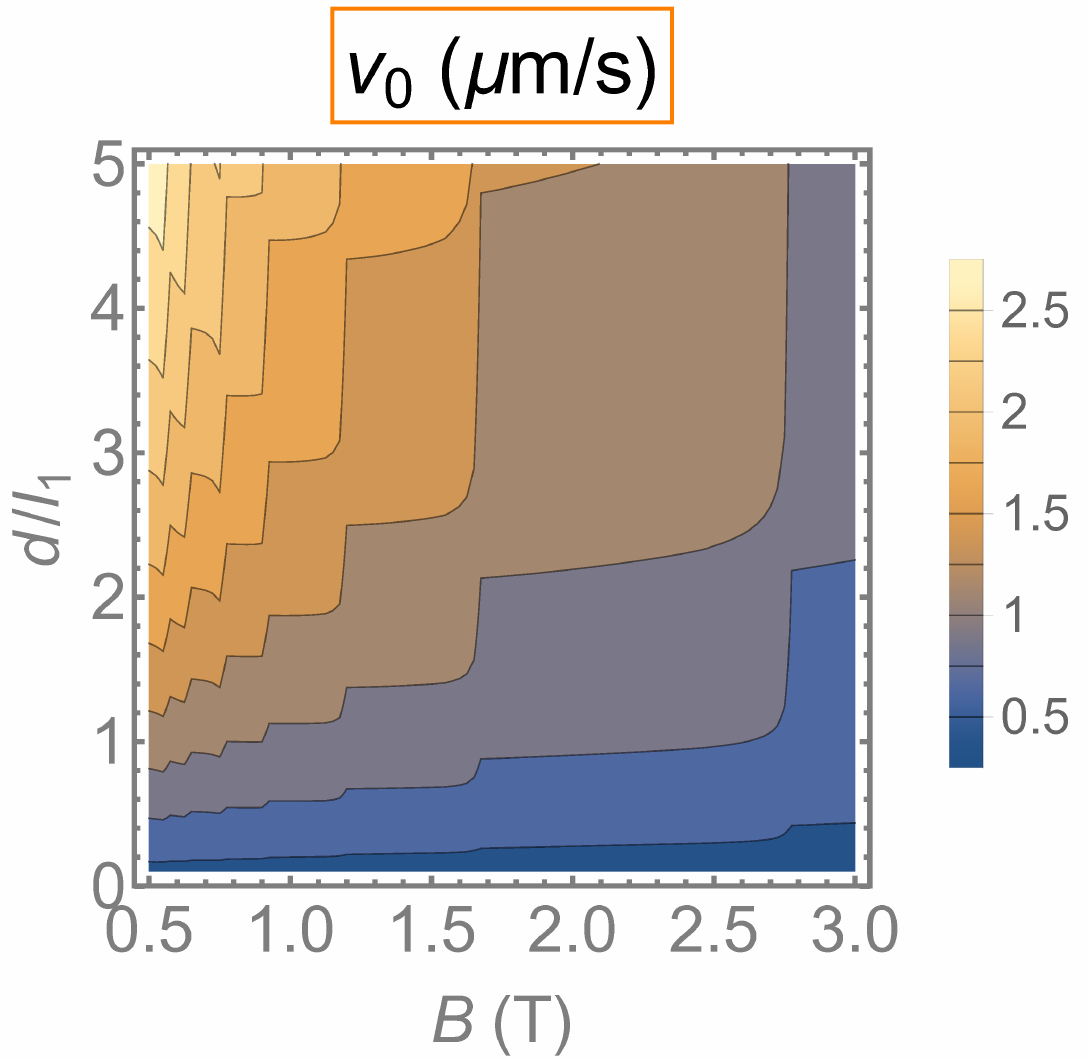}
(b)\includegraphics[scale=0.48]{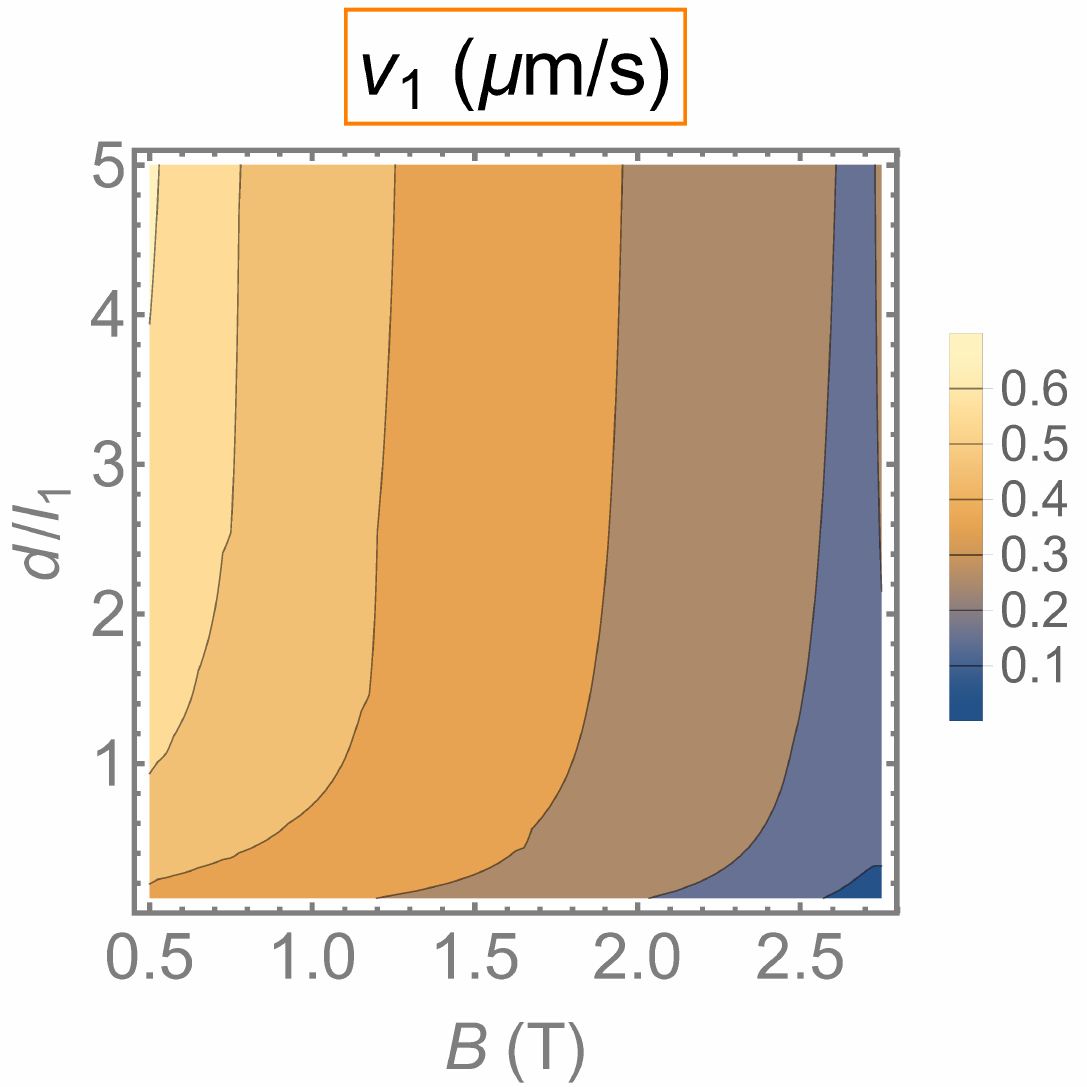}
(c)\includegraphics[scale=0.48]{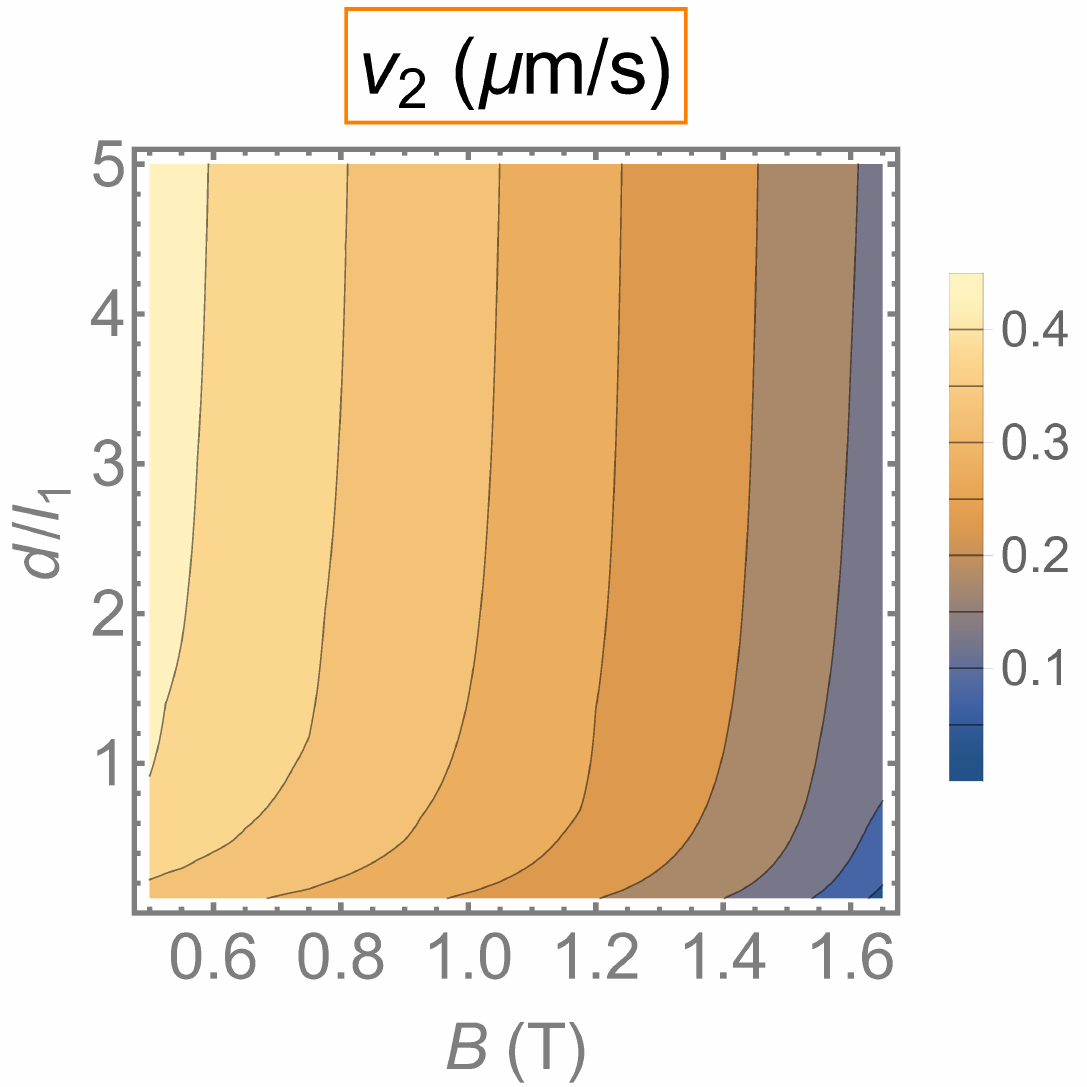}
\caption{\label{fig:sharp-2deg-velocities} Velocity of the three fastest EMPs supported by a sharp edge in a 2DEG, sorted from (a) to (c) in decreasing order of velocity. The plots show the dependence on the magnetic field (in Tesla) and on the distance $d$ of the top gate, in units $l_1\equiv l_B(B=1\mathrm{T})\approx 26\mathrm{nm}$.  For the plots, we used typical values of GaAs parameters, $\epsilon^*_S=8.7$, $m^*=0.063$, and $n_0=10^{11} \mathrm{cm}^{-2}$.
The steps as a function of $B$ occur when the Fermi energy crosses the bulk Landau levels: in these regions our model fails to account a finite real part of diagonal component of the conductivity and it is not applicable.}
\end{figure}

\begin{figure}
(a)\includegraphics[scale=0.55]{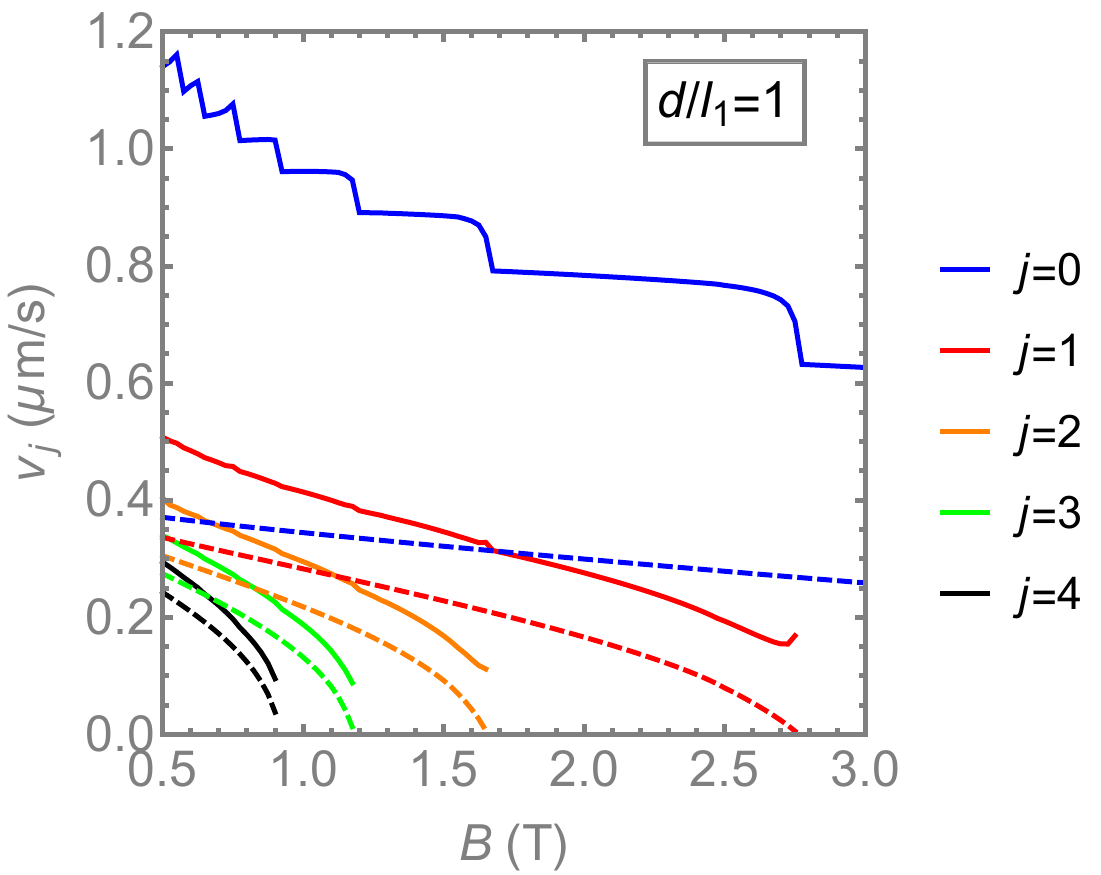}
(b)\includegraphics[scale=0.55]{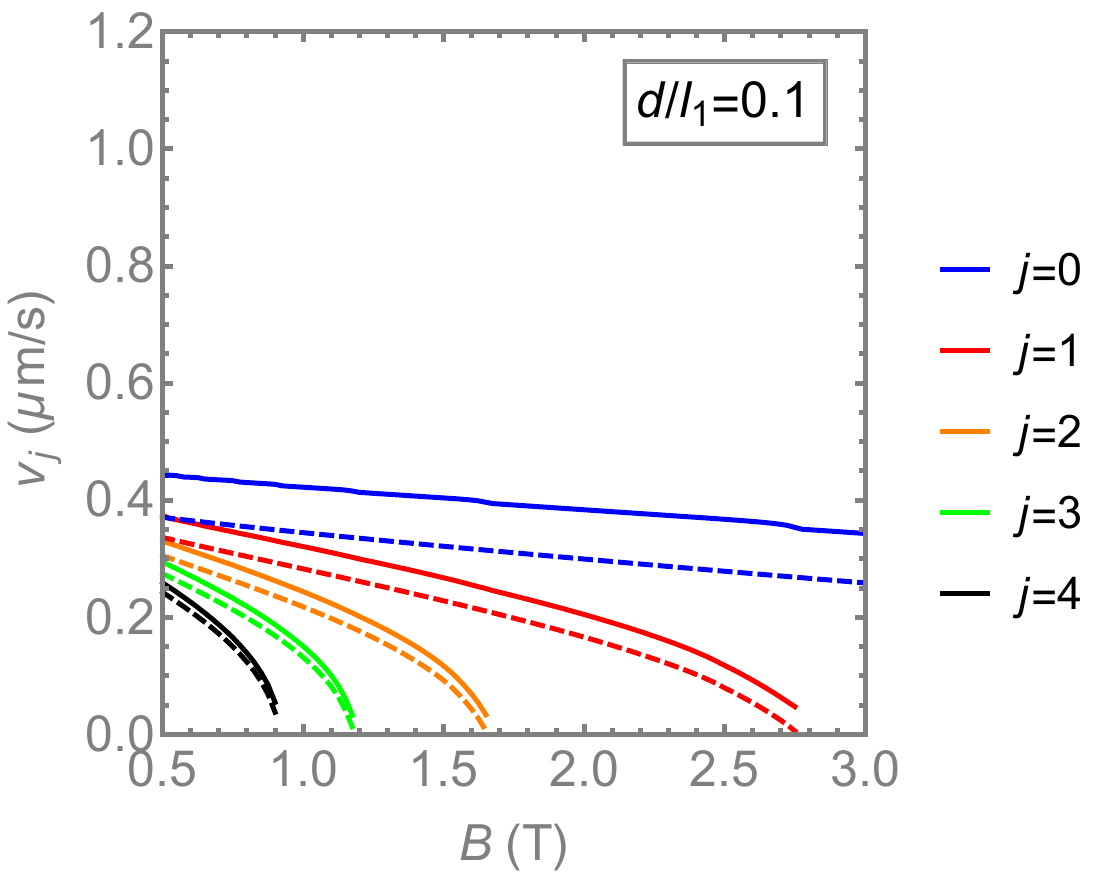}
\caption{\label{fig:sharp-2deg-velocities-B} Velocity of the EMPs supported by sharp edge in a 2DEG, as a function of magnetic field (in Tesla) for two different values of $d$ in units $l_1\equiv l_B(B=1\mathrm{T})\approx 26\mathrm{nm}$. In (a), we used $d/l_1=1$ and in (b), we used $d/l_1=0.1$. For the plots, we used typical values of GaAs parameters, $\epsilon^*_S=8.7$, $m^*=0.063$, $n_0=10^{11} \mathrm{cm}^{-2}$.
The solid lines represent the EMP velocities, while dashed lines represent the quantum contributions only. The quantum contributions are more relevant in (b). 
Our model is not applicable where the Fermi energy crosses bulk Landau levels, i.e. in the regions near the steep steps of the fastest EMP velocity in (a).}
\end{figure}

Fig. \ref{fig:sharp-2deg-velocities} shows the velocity of the first three modes as a function of magnetic field and distance to the top gate.
Consistent with the  free-space calculations in \citep{Mikhailov}, the velocities increase with $d$ and, while the velocity of the modes with $j\geq 1$ saturates to a finite value in the free-space limit, $d\rightarrow\infty$, the velocity of the fastest mode diverges.
When $d$ is comparable to $l_B$, the quantum contributions have a considerable impact on all the mode velocities; while, when the top gate is moved away from the 2DEG, the electrostatic contribution increases, in particular, in the fastest mode. In fact, it shows steps at magnetic fields corresponding to integer filling factor, where the matrix $\hat{\mu}$ changes size, as shown in Fig. \ref{fig:sharp-2deg-velocities-B}.

The behavior in the $x$-direction of the EMPs is easily found, but we do not show it here. As in the smooth edge case, the fastest mode is always the only monopole, while slower modes have a richer structure, not necessarily involving only $j$ nodes, depending on the details of eigenfunctions of $\hat{H}_0$ and the eigenvectors of $\hat{\mu}$.

Our model does not capture the modification in the plasmon velocities and their $x$-distribution due to scattering which are expected to occur when the Fermi energy crosses a bulk-LL.
Qualitatively, we expect an additional velocity term, dependent on $\mathrm{Re}(\sigma_{xx})$, as in \cite{Volkov}, to become relevant at small scattering timescale $(\omega_C\tau)_{\mathrm{min}}\approx1$, but we do not investigate this component further.


\subsection{\label{subsec:graphene}Sharp edges, graphene}

In graphene, the magnetic field dependent Hamiltonian $\hat{H}_B$, linearized in the vicinity of the Dirac points, takes the form \cite{CastroNeto, Schliemann} 
\begin{equation}
\hat{H}_B= \hbar \omega_c \left(\left(
\begin{array}{cc}
 0 & -\hat{a}_1^{\dagger} \\
 -\hat{a}_1 & 0 \\
\end{array}
\right)\oplus \left(
\begin{array}{cc}
 0 & \hat{a}_2 \\
 \hat{a}_2^{\dagger} & 0 \\
\end{array}
\right)\right),
\label{eq:MF-ham-graph}
\end{equation}
where  $\oplus$ is the direct sum and $\omega_c\equiv \sqrt{2} v_F/l_B\approx 26 \mathrm{meV}\sqrt{|B|/\mathrm{Tesla}}$ is the cyclotron frequency in graphene ($v_F\approx10^6$m/s is the Fermi velocity). The creation and annihilation operators labeled by $\sigma=(1,2)$ act only on the subspace of the valley near the $\sigma$th Dirac point; the $2\times 2$ matrices act on the $\sigma$th two-dimensional spinor
\begin{equation}
\psi_{\sigma} \equiv \left(
\begin{array}{c}
\phi^{\sigma}_a \\
\phi^{\sigma}_b \\
\end{array}
\right),
\label{eq:spinor-graph}
\end{equation}
where $a, b$ indicates the sublattice. Here, $\psi$ is intended to be a smooth envelope function in an effective-mass expansion for the wavefunction \cite{DiVincenzo-Mele}, and it can approximate the crystal wavefunction at lengthscales  greater than the Bohr radius $a_B\approx 0.5 \mathrm{\AA}$.

The eigenvalues of the Hamiltonian in Eq. (\ref{eq:MF-ham-graph}) are doubly-degenerate in the valley index $\sigma$ and they are
\begin{equation}
\epsilon_n=\pm \hbar \omega_c \sqrt{n},
\label{eq:MF-ham-eigenvalue-graph}
\end{equation}
with $n\in \mathbb{N}$ being the LL index.
Again, we neglect the Zeeman splitting correction \cite{CastroNeto}, and consider spin fully degenerate, leading to a additional factor $2$ in the EMP velocity as discussed in Sec. \ref{subsec:sharp-edges}.

Working again in the Landau gauge, the eigenvalues in Eq. (\ref{eq:MF-ham-eigenvalue-graph}) are infinitely degenerate in the momentum quantum number $k_y$. 
The termination of the honeycomb lattice breaks the translational invariance in the $x$-direction and therefore the degeneracy in $k_y$ is lifted.
The boundary conditions for low-energy excitations in graphene have been intensively studied with \cite{Akhmerov2, Brey-Fertig, Montambaux-graphene, Akhmerov4, Akhmerov3} and without applied magnetic field \cite{Akhmerov3, Akhmerov1, Falko}.
In particular, there are two fundamental classes of boundary conditions, zig-zag and armchair.
Zig-zag edges do not admix the valleys and the two Dirac cones can be treated separately, while armchair edges require combinations of two valleys and the valley degeneracy is lifted.

\begin{figure}
\includegraphics[scale=0.8]{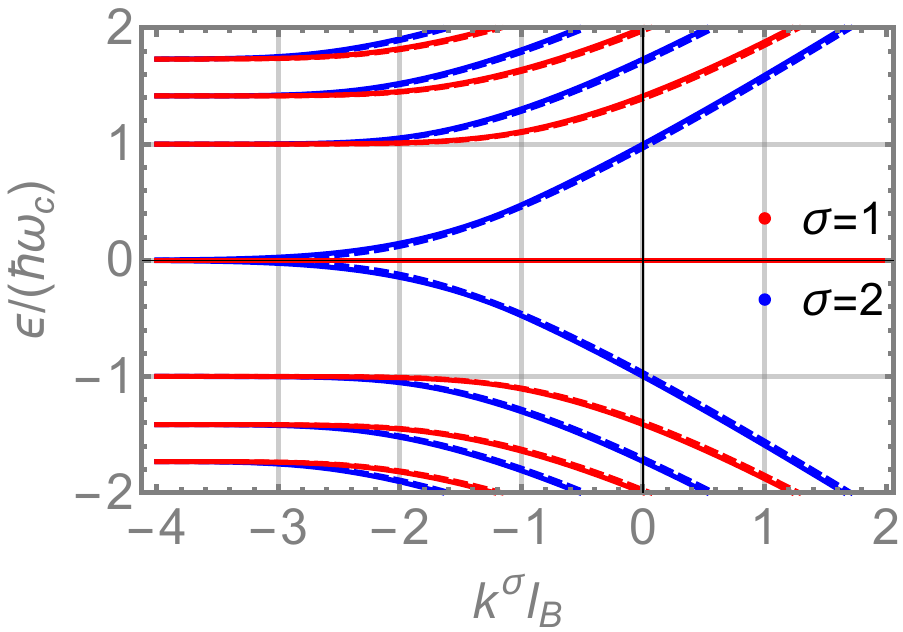}
\caption{\label{fig:band-graphene} 
Landau levels of monolayer graphene terminated by a zig-zag edge. Here, $\sigma$ is the valley (pseudospin) index. The energy levels are computed within the WKB approximation (solid line), and exactly (dashed lines). Static interactions have been neglected.}
\end{figure}

In the following, we focus only on former case, which is statistically more likely to occur \cite{Akhmerov1}, and we compute the band structure and the envelope functions $\psi$ with the WKB approach proposed in \cite{Montambaux-graphene}. 
A comparison between exact \cite{Akhmerov4} and WKB band structure for a monolayer graphene with zig-zag edges is shown in Fig. \ref{fig:band-graphene}.

Since the edges of graphene are terminated on the scale of the Bohr radius $w\approx a_B$, the sharp edge model applies well, and the validity of this approach was experimentally validated (without a top-gate) in \cite{Glattli, Glattli1, Glattli2}.

As the valleys are not mixed, the EMPs equations (\ref{eq:charge-sharp1}) and (\ref{eq:motion-sharp}) are modified simply by including an additional valley index $\sigma$. We then perform the substitutions $n\rightarrow (n,\sigma)$ and $m\rightarrow (m,\sigma')$ and use the 2-dimensional envelope function in Eq. (\ref{eq:spinor-graph}).
The dimension of the matrix $\hat{\mu}$ in Eq. (\ref{eq:velocity-sharp}) in graphene becomes then $2\nu_0-1$ instead of $\nu_0$ (the graphene filling factor $\nu_0$ is defined as the highest occupied bulk LL).

\begin{figure}
(a)\includegraphics[scale=0.48]{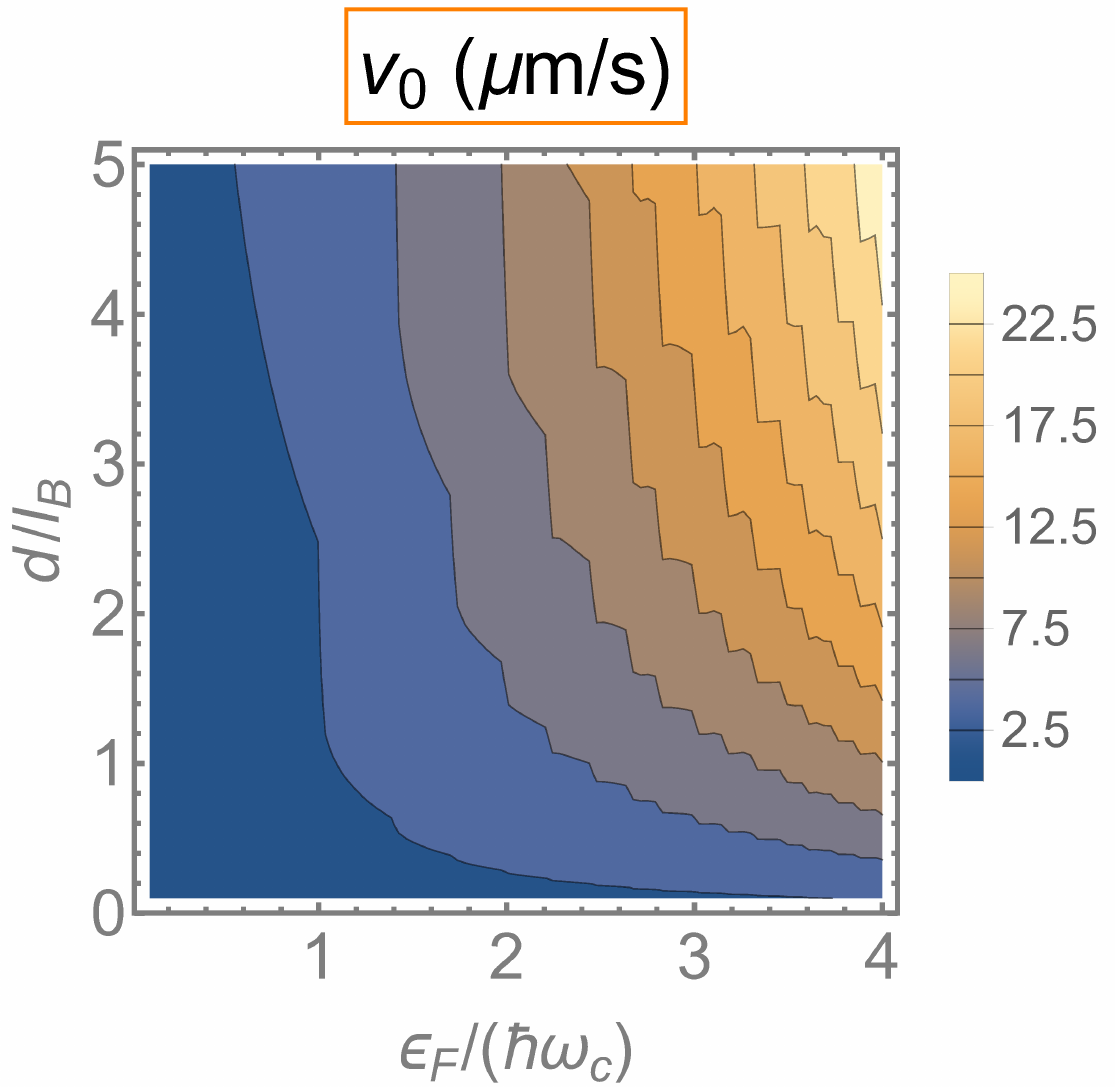}
(b)\includegraphics[scale=0.48]{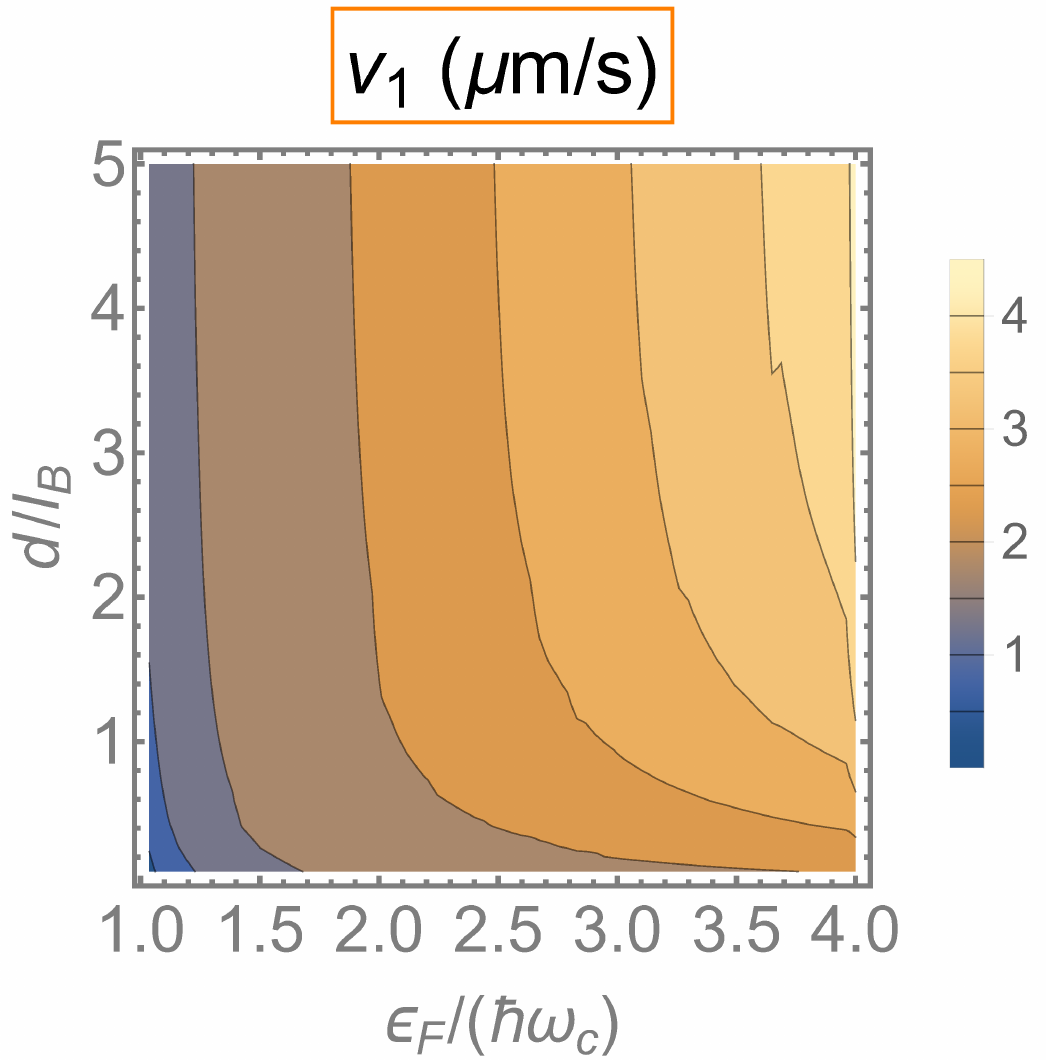}
(c)\includegraphics[scale=0.48]{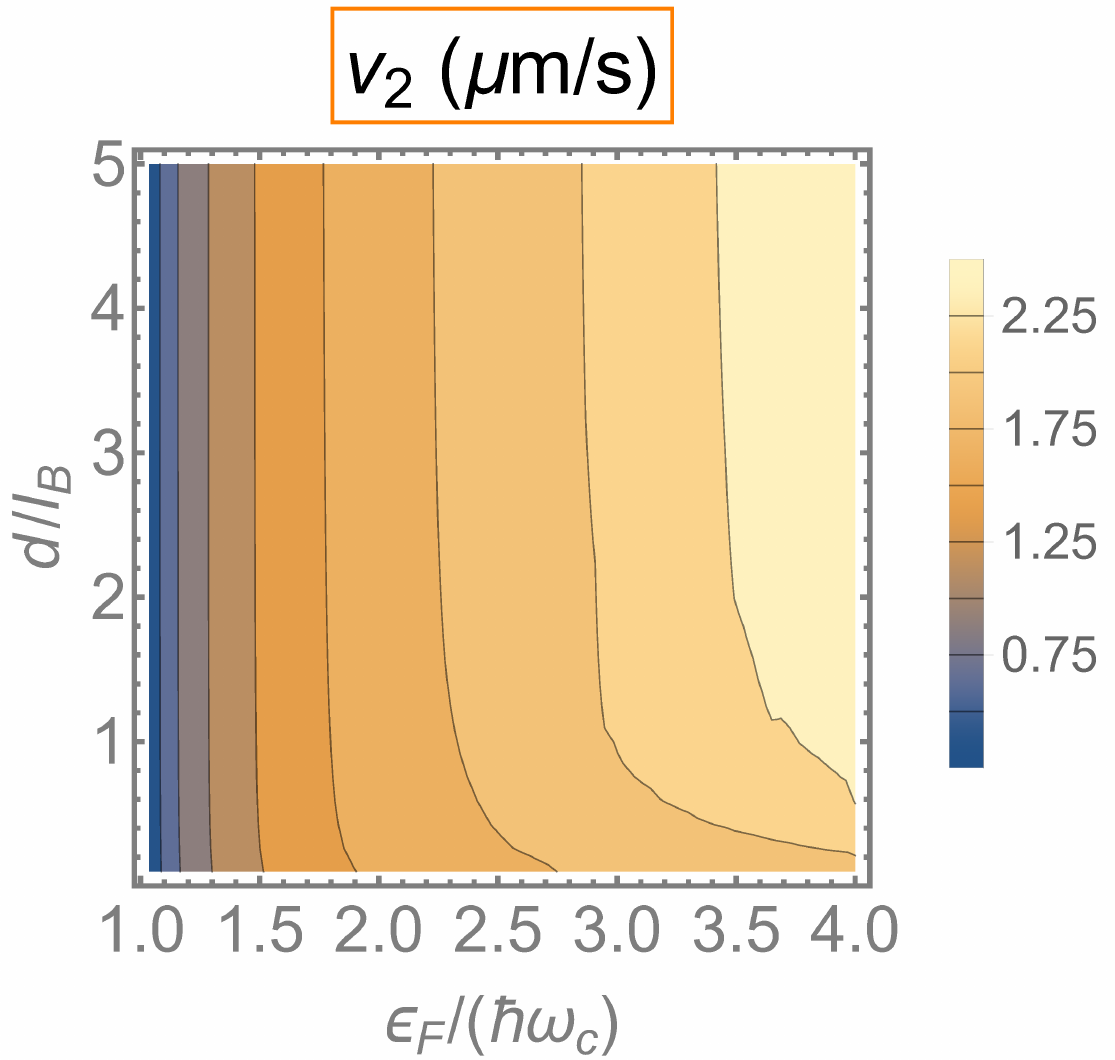}
\caption{\label{fig:sharp-graph-velocities} Velocity of the three fastest EMPs supported by a zig-zag edge of monolayer graphene, sorted from (a) to (c) in decreasing order of velocity. The plots show the dependence on the Fermi energy in units of the cyclotron energy $\epsilon_F/(\hbar \omega_c)$ and on the distance of the electrode in units of the magnetic length $d/l_B$. For the plot we used the typical value of the dielectric constant of silicon dioxide $\epsilon^*_S=3.9$.
When the Fermi energy crosses the bulk Landau levels, the fastest plasmon velocity has a step: in these regions our model fails to account for a finite real part of the diagonal component of the conductivity and it is not applicable.
}
\end{figure}

Figure \ref{fig:sharp-graph-velocities} shows the velocities of fastest EMP modes as a function of the Fermi energy $\epsilon_F$ (in units of the cyclotron energy $\hbar\omega_c$) and the distance of the top electrode $d$ (in units of the magnetic length $l_B$).
The plasmons becomes faster when the metal electrode is far away from the graphene sheet, consistent with the expected logarithmic divergence in wavevector in the free-space limit.
Moreover, the first mode is strongly influenced by Fermi energy variation and it shows the step-like behavior due to the QH plateaus.
However, as discussed for 2DEGs, our model is not applicable in the vicinity of these steps as there the diagonal components of the conductivity assume a finite value and an additional velocity term should be accounted for.

\begin{figure}
(a)\includegraphics[scale=0.55]{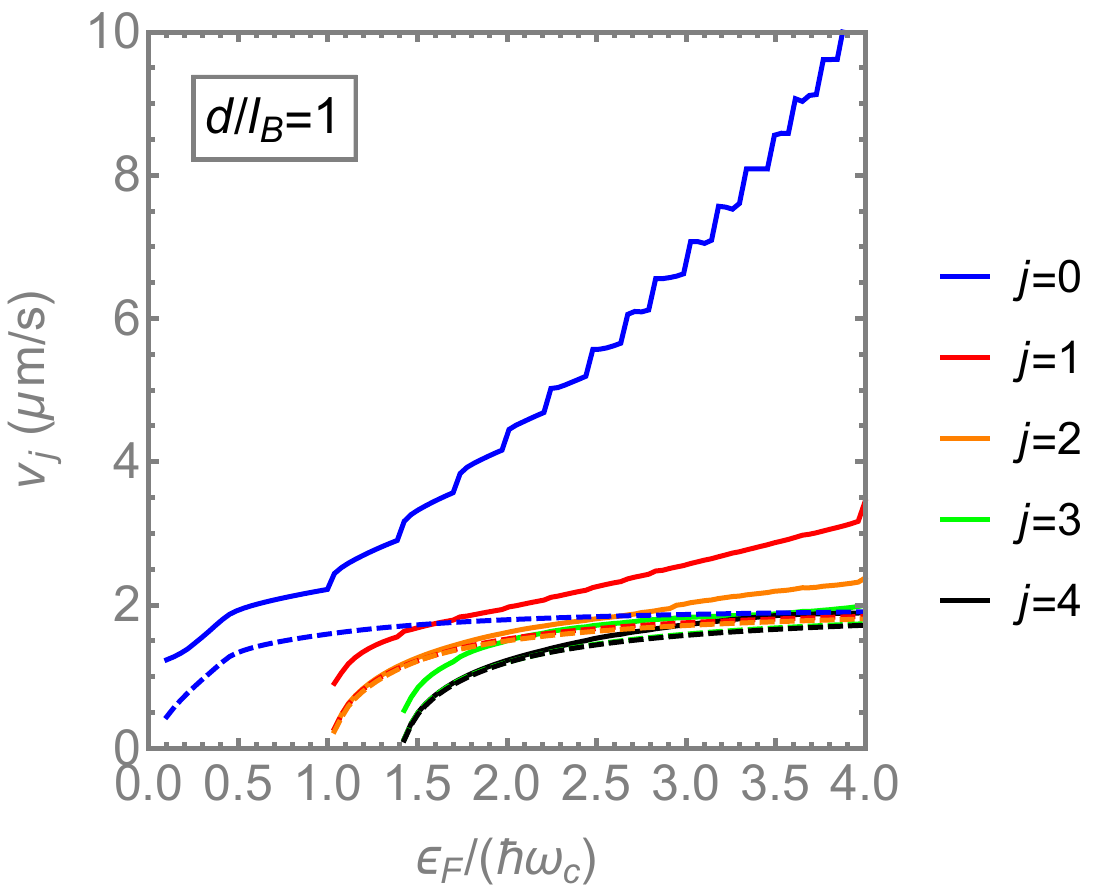}
(b)\includegraphics[scale=0.55]{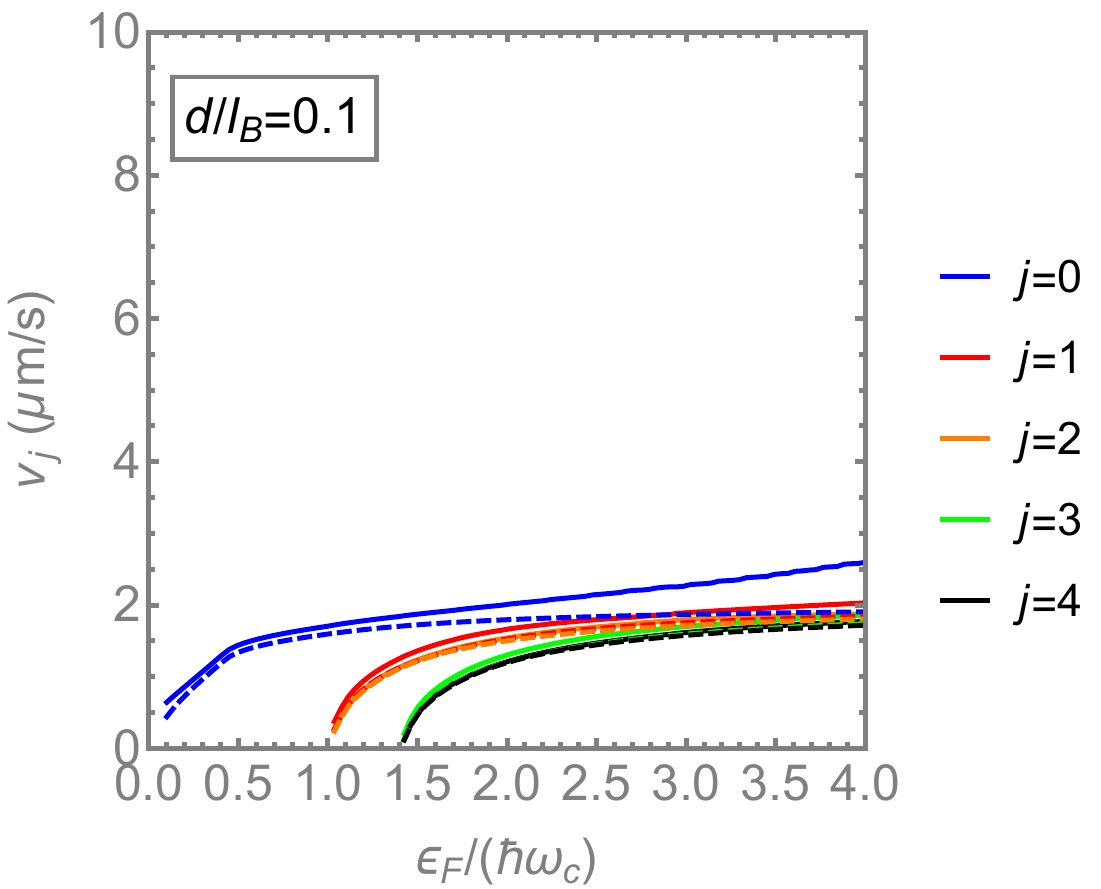}
\caption{\label{fig:sharp-graph-velocities-B} Velocity of the EMPs supported supported by a zig-zag edge of monolayer graphene, as a function of the Fermi energy (in units of the cyclotron energy) $\epsilon_F/(\hbar \omega_c)$ for two different values of $d/l_B$.  In (a), we used $d/l_B=1$ and in (b), we used $d/l_B=0.1$.
For the plots, we used the typical value of the dielectric constant of silicon dioxide $\epsilon^*_S=3.9$.
The solid lines represent the EMP velocities, while dashed lines represent the quantum contributions only.
Our model is not applicable where the Fermi energy crosses bulk Landau levels, i.e. in the regions in the vicinity of the steep steps of the fastest plasmon velocity in (a).}
\end{figure}

Note that due to the $\sqrt{n}$ dependence of the LL spacing typical in a Dirac-like material in a magnetic field, the LLs are more dense at high $\epsilon_F$ and the plateaus become shorter.
To examine in more detail the $\epsilon_F$ dependence and investigate the effect of the quantum velocities, we show in Fig. \ref{fig:sharp-graph-velocities-B} the mode velocities for two different $d$ values.
As for the 2DEG, quantum velocities play a fundamental role as the modes become slower; the fastest mode is dominated by the electrostatic contributions for $d/l_B\gtrsim 1$ and $\epsilon_F/(\hbar\omega_c)\gtrsim 1$.  


\subsection{\label{subsec:VD-RPA}Comparison with VD model}
A phenomenological  approach to model QH devices is proposed by Viola and DiVincenzo (VD) \cite{Viola-DiVincenzo}.
It is useful to work out explicitly the connection between our model and VD's.
VD considered a 1-dimensional line of charge $\rho(y,t)$ propagating at the boundary of the 2-dimensional material satisfying the capacitance and the Hall-effect current-voltage relations 
\begin{subequations}
\label{eq:BC-Laplace}
\begin{flalign}
\rho &=c (V_e-V),\\
\frac{\partial\rho}{\partial t} &=-\sigma_{xx}\frac{\partial V }{\partial x} -\sigma_{xy}\frac{\partial V}{\partial y}.
\end{flalign}
\end{subequations}
Here, $c$ is a phenomenological capacitance (per unit length) function describing the coupling with the electrodes. In the VD model, the electric potential $V$ in the 2-dimensional material obeys the Laplace equation \cite{Wick}
\begin{equation}
\nabla^2V=0,
\label{eq:Laplace}
\end{equation}
with boundary conditions obtained from Eq. (\ref{eq:BC-Laplace}).

In the QH regime, i.e.  $\sigma_{xx}\rightarrow 0$, the line charge obeys the linear partial differential equation
\begin{equation}
\frac{\partial \rho(y,t)}{\partial t} = \frac{\sigma_{xy}}{c} \frac{\partial \rho(y,t)}{\partial y}-\sigma_{xy}\frac{\partial V_e(y,t)}{\partial y}.
\label{eq:VD-charge}
\end{equation}
Comparing Eqs. (\ref{eq:motion-smooth-eigen-aj}) and (\ref{eq:VD-charge}) and  identifying $\sigma_{xy}/c$ with the velocity of the fastest plasmon $v_0$, we can notice that the VD model neglects all the slow modes and the details of the spatial distribution of the EMP in the interior of the material.
This approach can be justified when $v_0$ is much greater than all the other velocities, such that slower modes are less coupled to the electrode and strongly damped compared to the fastest, which dominates the response of the device.

Note also that since $v_0$ includes both electrostatic and quantum contributions in our model, we can now quantitatively define the effect of the quantum capacitance.
From Sec. \ref{subsec:sharp-edges}, the quantum capacitance plays an important role for sharp edges, especially when the top gate is very near the 2-dimensional material. In this case, the inverse effective capacitance is obtained by diagonalizing the matrix $\hat{\mu}$, defined for 2DEGs in Eq. (\ref{eq:velocity-sharp}); it reduces simply to the sum of a quantum and a geometric inverse capacitance only when $\nu_0=1$.

\section{\label{sec:application} Application: 3-Terminal gyrator}

We now use the EMP model described in the Sec. \ref{sec:model}  to compute the response of a QH bar of perimeter $L_y$ capacitively coupled to $N$-electrodes.
For simplicity, we only consider devices with a smooth-shaped boundary, whose local radius of curvature is much greater than $\mathrm{max}(l_B,w)$. In this case, the EMP equations obtained for the straight line geometry still hold, and $y$ parametrizes the position on the boundary of the device. This is consistent with the applied periodic boundary conditions to the wavefunctions.
Note that, in graphene, we are neglecting the valley mixing phenomenon that inevitably occurs in closed devices.

At first, we neglect damping and set $\eta=0$.
We assume that the coupling between each of the electrodes and the Hall bar is constant in space or, in other words, that the velocity of EMP stays constant around the perimeter of the device. This approximation holds when the gap between different electrodes is much shorter than the wavelength of the plasmon.

To model a device, we then consider an external drive
\begin{equation}
V_e(x,y,t)=\sum_{n=1}^N V_n(t)\left(\theta(y-y_n)-\theta(y-y_n-L_n)\right),
\label{eq:voltage-external}
\end{equation}
where $V_n(t)$ is the time dependent drive applied at the $n$th electrode, and $y_n$ and $L_n$ define respectively its initial coordinate and its length.
Again, we assumed that the top electrode extends inside the material far further than the EMP, and so the external drive can be considered constant in $x$ (the direction normal to the boundary).

The current flowing in the external electrodes depends on the discontinuity of normal derivative of the displacement field at position $z=d$, and it is related to the EMP charge $\rho_1$ by
\begin{equation}
I_{n}(t)= \epsilon_S \int_{S_n} d \overline{r}\int_{\mathbb{R}^2} d \overline{r}' \frac{\partial}{\partial z}G(\overline{r},\overline{r}',z\rightarrow d)\frac{\partial\rho_1(\overline{r}',t)}{\partial t}.
\label{eq:current-displacement}
\end{equation}
Here, $S_n$  is the surface of the $nth$ electrode, and $G$ is the electrostatic Green's function of the 3-dimensional Poisson operator.
Note that $G$ should account for the geometry of the device, and in particular for the physical gap between electrodes.
For simplicity, we neglect these gaps in computing $G$, and we use the Green's function shown in Eq. (\ref{eq:Greens-top-gate-3d}). This choice is consistent with considering the velocity of the EMP constant along the perimeter and it has the same limits of applicability. 

\begin{figure}
\includegraphics[scale=0.3]{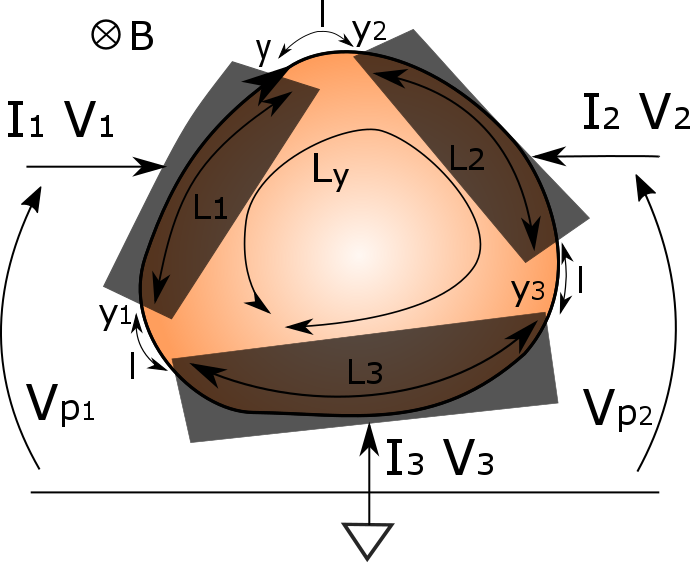}
\caption{Three-terminal gyrator. Three electrodes of length $L_n$ are capacitively coupled to a Hall bar of perimeter $L_y$, whose boundary is parametrized by $y$. The starting position of the electrodes in the $y$-direction is labeled by $y_n$ and they are separated by gaps of length $l$ with the nearest ones. These gaps are assumed to be equal and small compared to $L_n$. The device behaves as a two-port gyrator when the voltage of two electrodes is measured with respect to the last one, which acts as a common ground. The convention of currents and voltages is shown in the plot.}
\label{fig:3-terminals-gyrator}
\end{figure}

We now  focus on the symmetric 3-terminal QH gyrator shown in Fig. \ref{fig:3-terminals-gyrator}.
In the VD model, this device is predicted to be a perfect gyrator at frequency 
\begin{equation}
\omega_n=\frac{\pi \sigma_{xy}}{c L_3 }(2n+1), \ \ \  n\in \mathbb{N},
\label{eq:gration-freq}
\end{equation}
and to have a very low impedance when $L_3=2L_1$ \cite{Bosco}.

Working in the frequency domain $t\rightarrow\omega$, the current $I_n$ flowing in the external $n$th electrode can be written in the matrix form
\begin{equation}
I_n(\omega)=\sum_{m=1}^3 Y_{nm}(\omega)V_m(\omega).
\end{equation}
The admittance matrix elements are derived in Appendix \ref{app:admittance} and they are given by
\begin{subequations}
\label{eq:admittance}
\begin{align}
\begin{split}
Y_{n n}(\omega)=& 2 i \sum _j q_j \csc  \left(\frac{\omega  L_y}{2 v_j}\right) \times\\
 &\left(\cos  \left(\frac{\omega ( L_y-2 L_n )}{2 v_j}\right)-\cos  \left(\frac{\omega L_y}{2 v_j}\right)\right),
 \end{split}\\
 Y_{n m}(\omega)=&\sum _j q_j f_j^{n m}(\omega)e^{i\omega L_y/v_j} , \ \ \  n<m,\\
  Y_{n m}(\omega)=&\sum _j q_j f_j^{n m} (\omega) , \ \ \  n>m,
 \end{align}
 \end{subequations}
with
\begin{multline}
f_j^{n m}(\omega)= \left(1+ i \cot\left( \frac{\omega L_y}{2 v_j}\right) \right)e^{i\omega(y_n-y_m-L_m)/v_j}\times \\
\left(1- e^{i\omega L_n/v_j}\right)\left(1- e^{i\omega L_m/v_j}\right),
\end{multline}
and 
\begin{equation}
q_j= -\frac{\sigma_{xy}}{2} \left\lbrace 
\begin{array}{ll}
M_{0j}^2  & \mathrm{for \ \ smooth \ \  edges}\\
\left(\sum_i M_{i j}\right)^2/\nu_0 & \mathrm{for \ \ sharp \ \ edges}.
\end{array}
\right. 
\end{equation}
For smooth and sharp edges, the transformation $\hat{M}$ diagonalizes the velocity matrix $\hat{\mu}$, defined respectively by Eqs. (\ref{eq:electrostatic-mu}) and (\ref{eq:velocity-sharp}), and the off-diagonal conductivity $\sigma_{xy}$ is intended to be respectively the classical and quantum HE conductivity.

From Eq. (\ref{eq:admittance}), it appears that the total admittance of the device is obtained by summing all the contributions of the single EMPs; this implies that an incoming signal can propagate through a set of \textit{parallel} chiral paths, each  of which is associated to a different plasmon mode.

Note that the column elements of the admittance matrix do not sum to zero, and this leads to a violation of the Kirchhoff's current law. This violation can be traced back to the approximation on the electrostatic Green's function $G$: by neglecting the changes in $G$ due to the gaps $l$ between the electrodes, additional (unphysical) currents can flow there.
If the gaps are much shorter than the terminals, $l\ll L_n$, these currents are small, but they should be accounted for to recover current conservation.

One can circumvent this problem in the analysis of the port response of the device by conveniently choosing an AC reference potential. In fact, Eq. (\ref{eq:motion-smooth-eigen-aj}) is invariant under the transformation $V_e(y,t)\rightarrow V_e(y,t)+V_b(t)$, and so we set the potential $V_b(t)$ to be a linear function of the applied voltages $V_n(t)$ such that the unwanted currents sum to zero.
Then, defining the port voltages and currents
\begin{subequations}
\begin{align}
V_{p_1}&=V_1-V_3, &&  & I_{p_1}&=I_1,\\
V_{p_2}&=V_2-V_3, &&  & I_{p_2}&=I_2,
\end{align}
\label{eq:ports-gyrator}
\end{subequations}
we find the 2x2 port admittance matrix $Y_p$ and, using the standard relation \cite{Pozar}
\begin{equation}
S=-\left( Y_p+\frac{1}{Z_0}\mathcal{I}\right)^{-1}\left( Y_p-\frac{1}{Z_0}\mathcal{I}\right),
\label{eq:S-matrix-Pozar}
\end{equation}
we evaluate the $S$ parameters of the device.
Here, $Z_0$ is the characteristic impedance of the external circuit, typically $50\Omega$, and $\mathcal{I}$ is the identity matrix.

To quantify the gyrating properties of the device, we introduce the parameter \cite{Viola-DiVincenzo, Bosco}
\begin{equation}
\label{eq:Deltapar}
\Delta \equiv\frac{1}{2}\lvert S_{21}-S_{12}\rvert \leq 1,
\end{equation}
where the equality is attained only for a perfect gyrator, with the $S$-matrix given in Eq. (\ref{eq:s-g-matrix}).

Let us first assume that only the fastest plasmon is excited and it is not damped and define, in analogy with \cite{Bosco}, the dimensionless frequency and impedance mismatch 
\begin{subequations}
\begin{flalign}
\Omega & \equiv \frac{\omega L_1}{v_0},
\label{eq:dimensionless freq} \\
\alpha & \equiv -4q_0Z_0.
\end{flalign}
\end{subequations}

\begin{figure}
\includegraphics[scale=0.8]{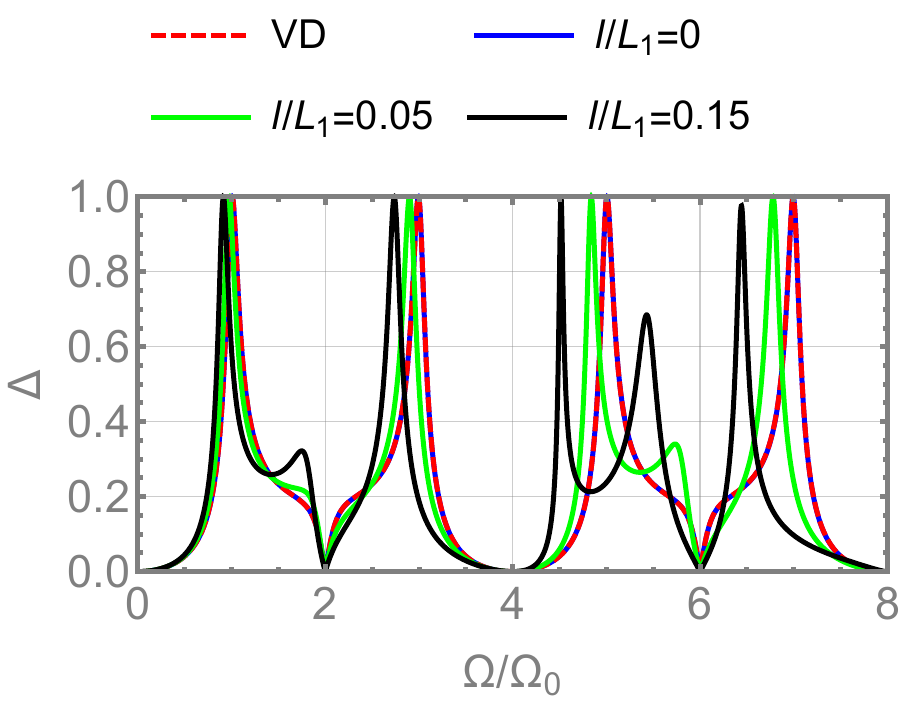}
\caption{Single mode response. We plot the dependence of the $\Delta$ parameter, defined in Eq. (\ref{eq:Deltapar}), on the dimensionless frequency $\Omega$, defined in Eq. (\ref{eq:dimensionless freq}). We normalize  $\Omega$  over the first gyration frequency $\Omega_0=\pi L_1/L_3$ and consider different values of the gap $l$ between electrodes. For the plot we used $\alpha=0.2$ and $L_3/L_1=2$.}
\label{fig:delta-phen-comparison}
\end{figure}

We can now compare our results with the one obtained with the VD model, after identifying $c=\sigma_{xy}/v_0$.
In Fig. \ref{fig:delta-phen-comparison},  we show how $\Delta$ varies with the dimensionless frequency, normalized over the first gyration frequency $\Omega_0=\pi L_1/L_3$ in the case $L_3=2L_1$. Our model, in the one mode approximation, exactly coincides with VD when there is no gap between electrodes. 
This is expected as in the VD model, in the QH regime, the charge is transmitted instantaneously between different electrodes, as its velocity in the gaps is $\propto 1/c \rightarrow \infty$; hence the gaps do not play any role \cite{Viola-DiVincenzo}.
In our model, however, the charge propagates along the perimeter at constant velocity, and thus finite gaps change the response, both shifting the gyration frequency and adding new features in $\Delta$.
Note that as $\alpha$ decreases, the band of good gyration becomes narrower \cite{Bosco}. For this reason, to make the additional features of our model more visible, we use in all the plots a rather high value of $\alpha=0.2$, i.e. $Z_0\gg50\Omega$, which would require an external impedance matching circuit. 

\begin{figure}
\includegraphics[scale=0.3]{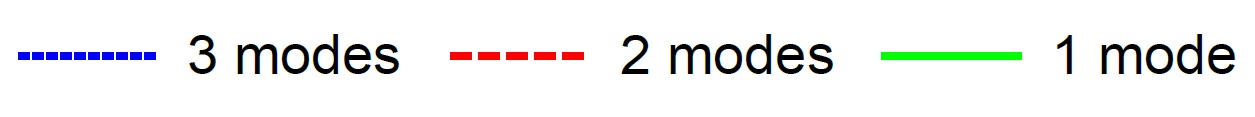}
(a)\includegraphics[scale=0.6]{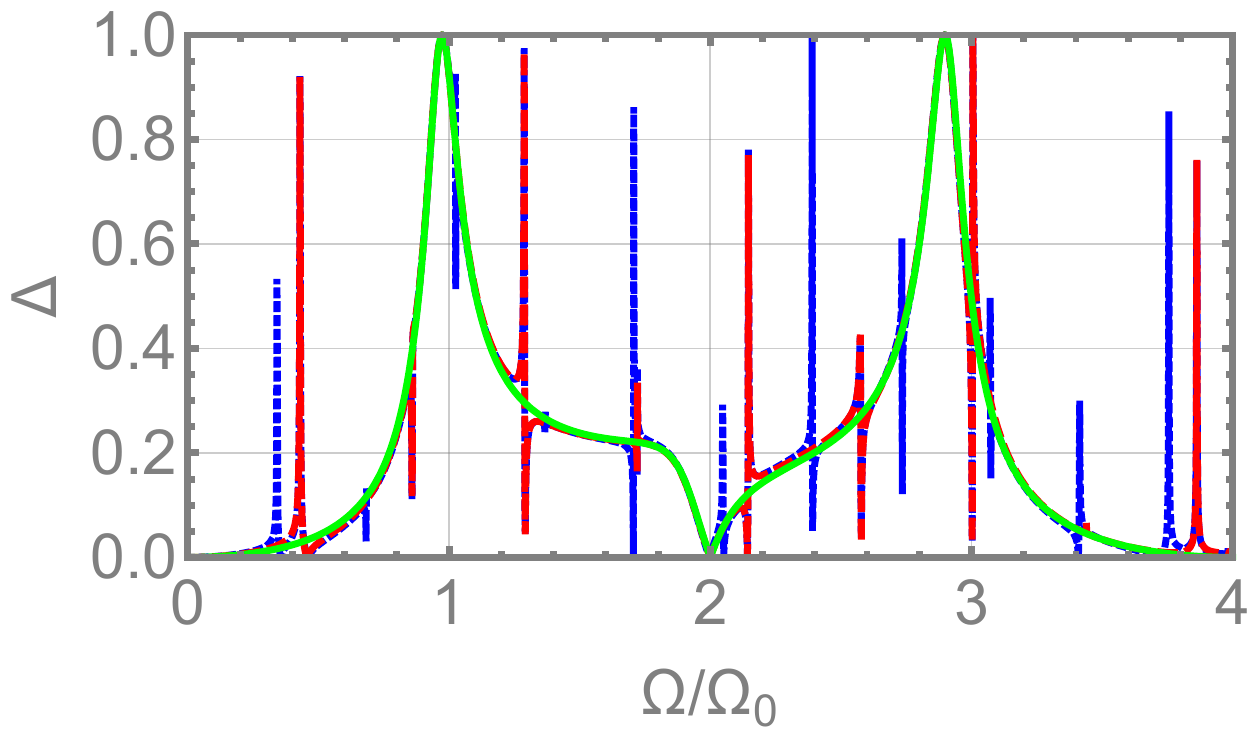}
(b)\includegraphics[scale=0.6]{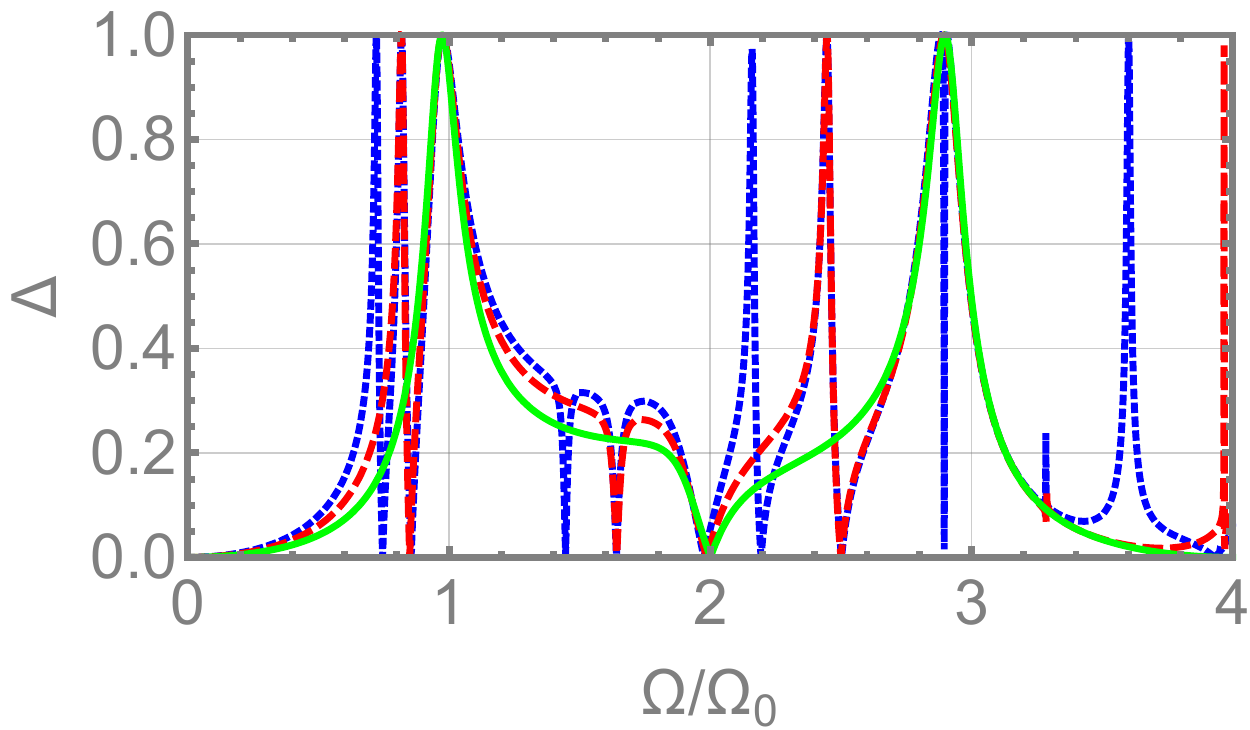}
(c)\includegraphics[scale=0.6]{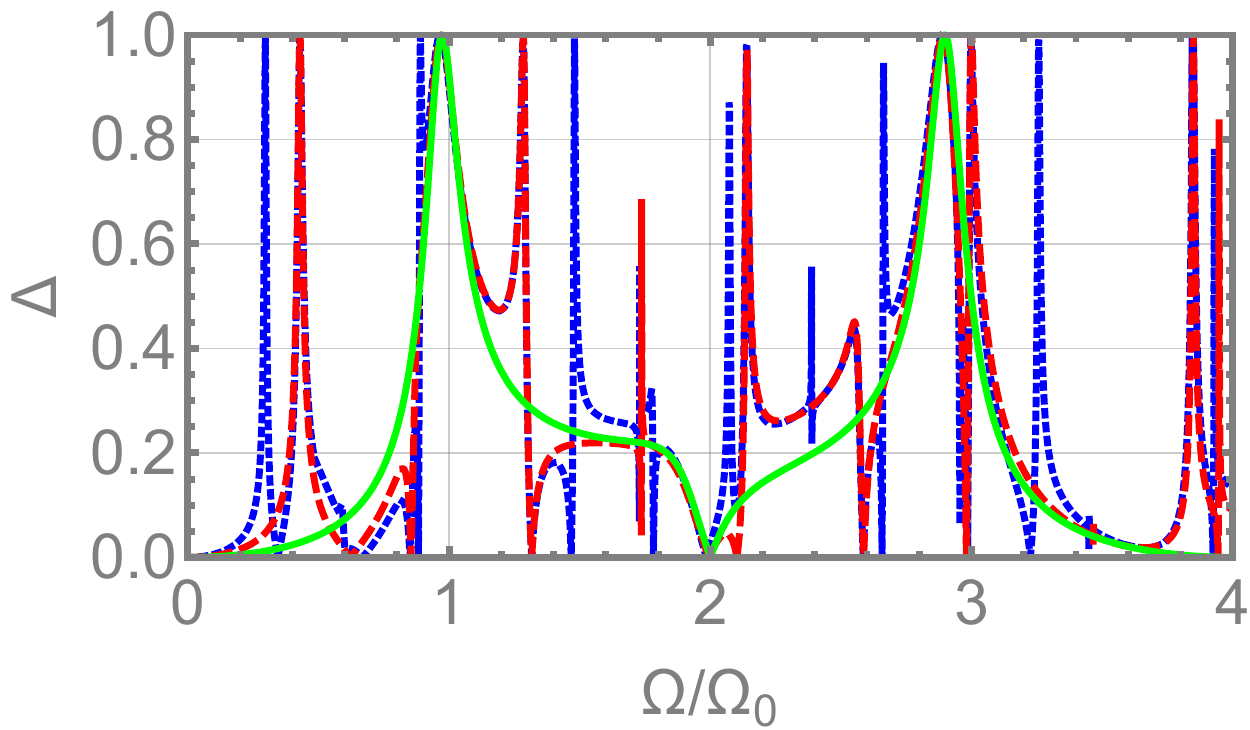}
\caption{Response including multiple modes. We plot the dependence of $\Delta$, defined in Eq. (\ref{eq:Deltapar}), on the dimensionless frequency $\Omega$, defined in Eq. (\ref{eq:dimensionless freq}). We normalize  $\Omega$   over the first gyration frequency $\Omega_0=\pi L_1/L_3$ and we include up to 3 modes in the response. For the plot, we used $\alpha=0.2$, $L_3/L_1=2$ and $l/L_1=0.05$. 
In (a) and (b), we consider the EMPs in a 2DEG with sharp edges (with the same parameters used in Fig. \ref{fig:sharp-2deg-velocities}), at $B=0.5$T and with $d/l_1=1$ and $d/l_1=0.1$ respectively.
In (c), we consider the EMPs in a 2DEG with smooth edges, with $d/w=0.1$. 
All the resonant structures go up to a value very close to 1 (or to 0), but the plots do not have enough resolution to show this behavior.}
\label{fig:delta-modes}
\end{figure}

We now investigate the effect of the slower modes.
In Fig. \ref{fig:delta-modes}, we show how $\Delta$ is affected by accounting for additional EMP modes in different situations. The slower modes add some resonant peaks in the response, whose position and broadening  (in frequency) strongly depends on the ratio $v_j/v_0$ and $q_j/q_0$. 
Note that all these sharp features go very close to the extremal values (0 or 1), but the resolution of the plot is not high enough to capture this behavior for the narrowest ones.

In particular, when the gaps are very close to each others, $l\rightarrow0$, one can easily distinguish two classes of resonance features caused by the $j$th mode. The first class includes all the asymmetric peaks that reach the limiting value $\Delta=1$ and they are centered at the normalized frequencies $\Omega^a_j=\pi v_j/v_0(n+1/2)$, with $ n\in \mathbb{N}$. The second class includes the remaining downward peaks, that are  centered at the frequencies $\Omega_j=2\Omega^a_j$.

The structure of the additional resonances very much resembles a Fano line-shape \cite{Fano}: electromagnetic waves with different resonant frequencies are known to interfere, leading to asymmetric peaks. In our case, the sharp resonances (due to the slow plasmonic modes) are superimposed upon a smoothly varying background (due to the fastest mode) in a way that seems qualitatively to agree with the Fano-resonance model developed for photonic crystals in \cite{Fan}.

Comparing Fig. \ref{fig:delta-modes} (a) and (b), one sees that when the fastest mode is strongly dominant, e.g. when the top gate is far from the 2DEG, the peaks are narrower and closer in frequency.
In graphene, we can observe the same behavior, but we do not report it here.
For smooth edges, the different definition of $q_j$ leads to some differences in the response, in particular, one can can observe that the peaks are closer to each other, but broader.

\begin{figure}
\includegraphics[scale=0.3]{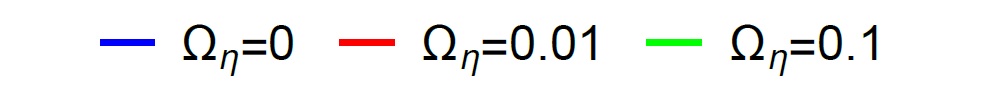}
(a)\includegraphics[scale=0.48]{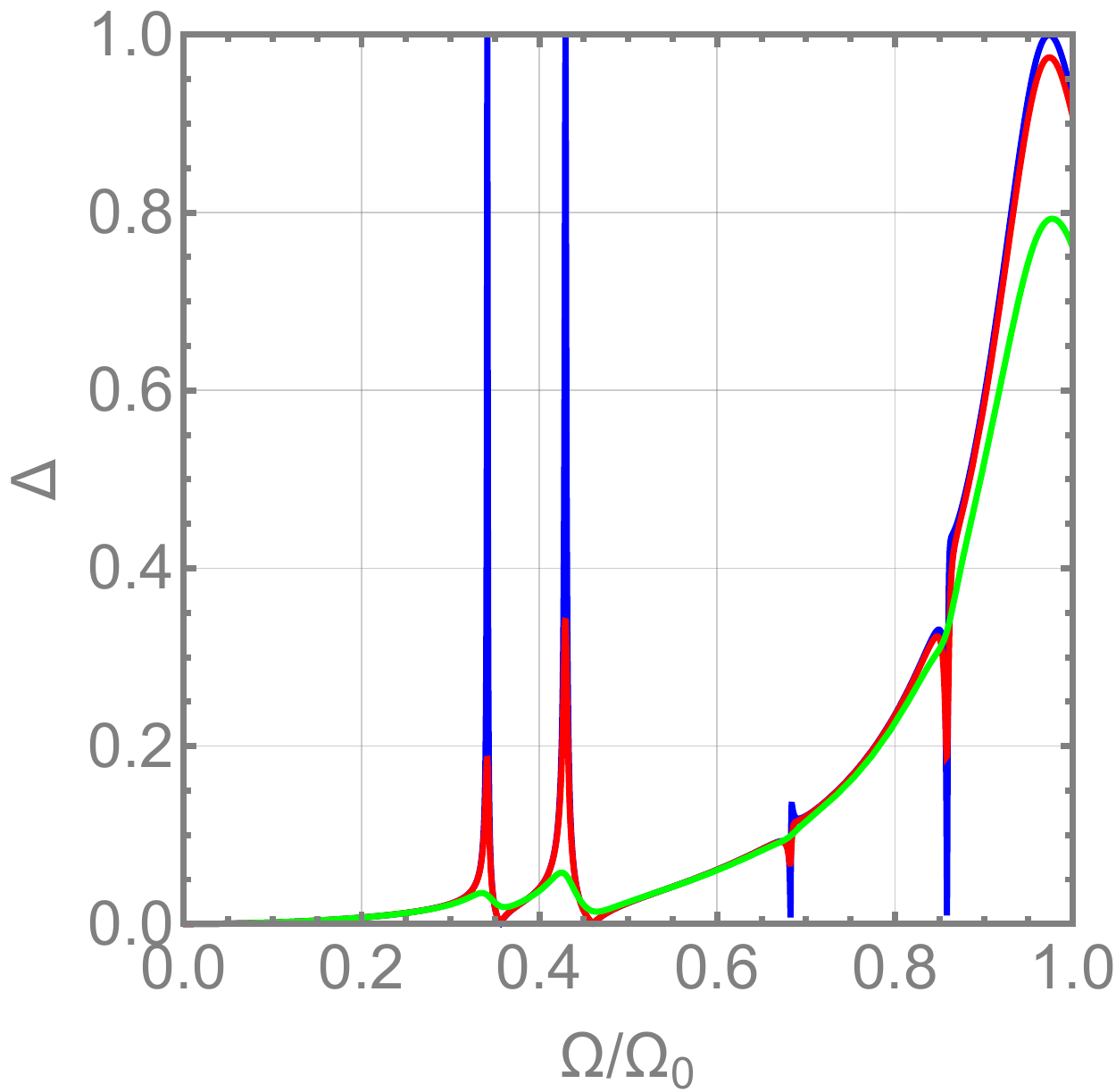}
(b)\includegraphics[scale=0.48]{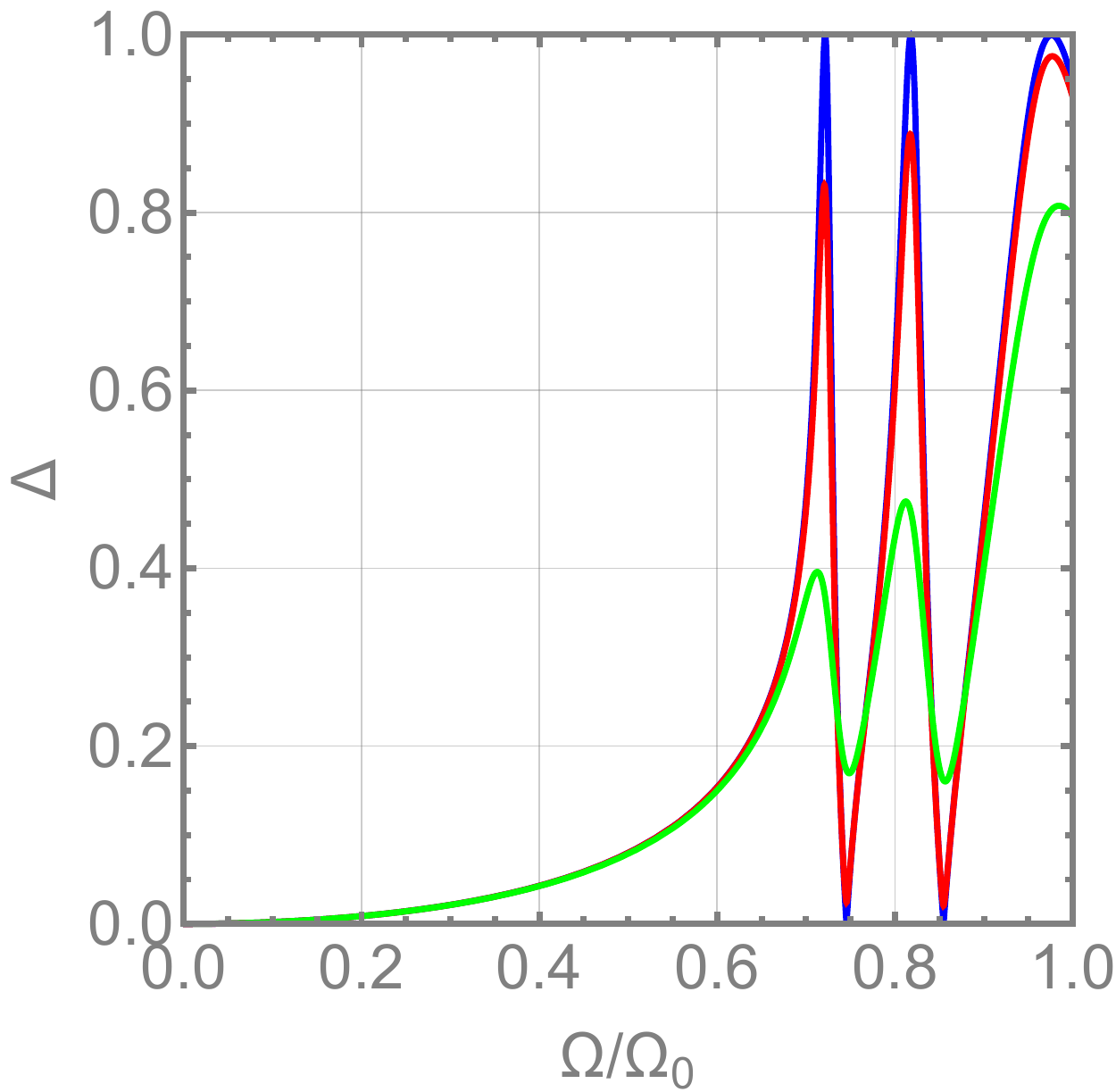}
(c)\includegraphics[scale=0.48]{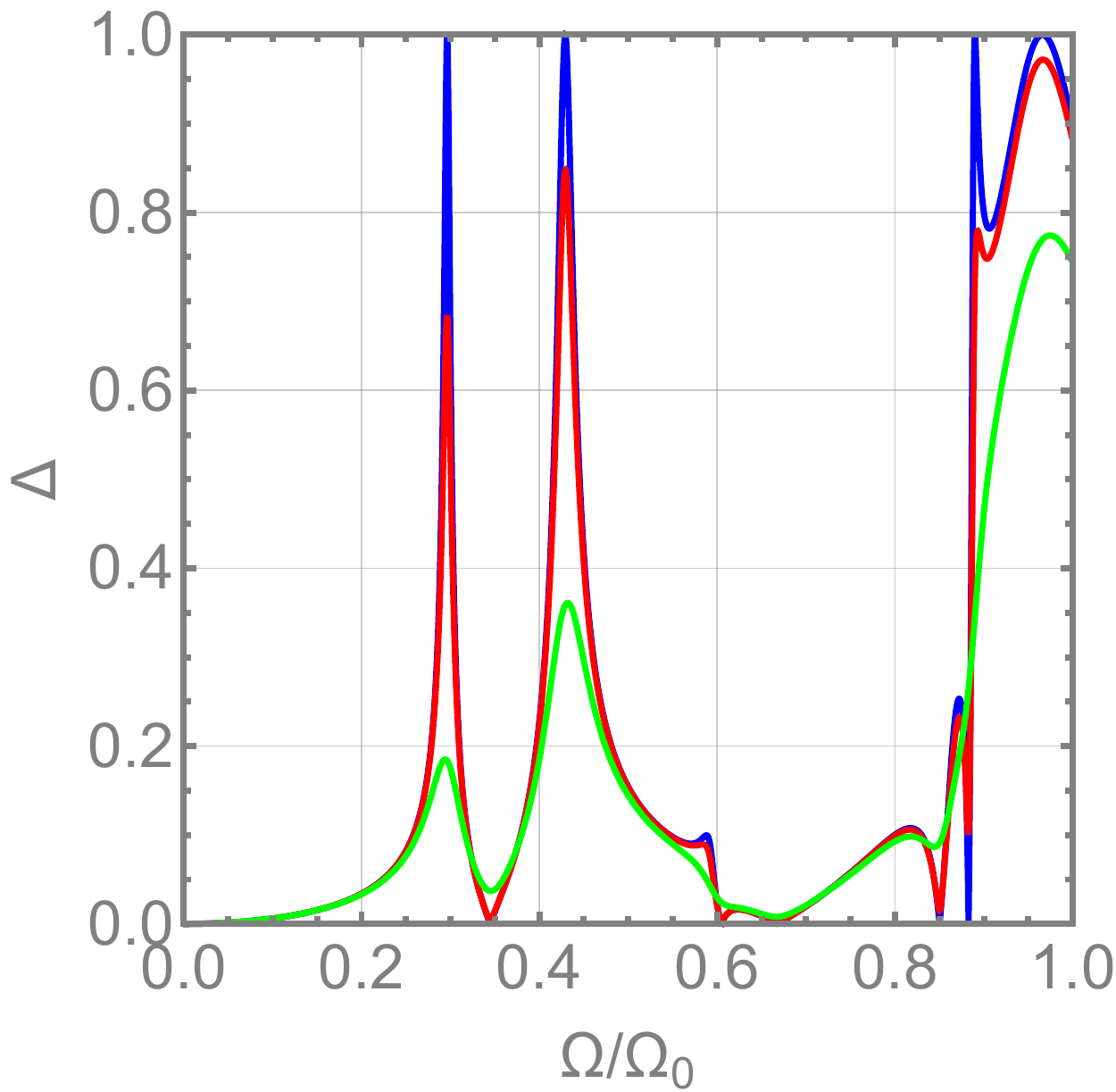}
\caption{Effect of a damping rate.  We plot the dependence of $\Delta$, defined in Eq. (\ref{eq:Deltapar}), on the dimensionless frequency $\Omega$, defined in Eq. (\ref{eq:dimensionless freq}). We normalize  $\Omega$ the first gyration frequency $\Omega_0=\pi L_1/L_3$ and use different damping rates $\Omega_{\eta}\equiv \eta L_y/v_0$. Here, the plots are obtained in the same way as the ones in Fig. \ref{fig:delta-modes}, and we zoomed in a narrower frequency range.
In (a) and (b), we consider the EMPs in a 2DEG with sharp edges (with the same parameters used in Fig. \ref{fig:sharp-2deg-velocities}), at $B=0.5$T and with $d/l_1=1$ and $d/l_1=0.1$ respectively.
In (c), we consider the EMPs in a 2DEG with smooth edges, with $d/w=0.1$. }
\label{fig:delta-damp}
\end{figure}

Let us now include a phenomenological damping rate $\eta$: Eq. (\ref{eq:admittance}) then is simply modified by the substitution $\omega\rightarrow \omega+i\eta$.
As the EMP modes decay with a lengthscale $v_j/\eta$, the peaks corresponding to the slower modes will be more affected by damping and they will be more difficult to observe.
We introduce the parameter $\Omega_{\eta}\equiv \eta L_y/v_0$, which characterizes the inverse of the number of laps around the device that the fastest EMP can perform before it is damped.
In Fig. \ref{fig:delta-damp},  we show how three responses are affected by the damping in the same situations as in Fig. \ref{fig:delta-modes}. As expected, the damping decreases the resonance peak heights, especially the ones due to slower EMPs.

The fact that the fastest mode dominates the response even for reasonably small damping can be useful in actual devices, where additional modes can cause distortion in the signal.


\section{Conclusions and outlook}

Driven by the recent research in the field of non-reciprocal devices exploiting the QH effect, we develop here a microscopic theory, based on linear response theory and RPA, to describe their response.

The device model is based on an  analysis of driven, chirally propagating EMPs supported by smooth and sharp edges.
Although the model offers several new insight into the device response, it still lacks a quantitative analysis of dissipation, which is now accounted for only through a phenomenological timescale.
An exhaustive treatment of the EMP decay is essential to better characterize and engineer QH devices; to this aim, good starting points would be the qualitative discussions on the possible loss mechanisms in \cite{Mikhailov} and the hydrodynamic analysis in \cite{Glazman,JohnsonVignale}.

Moreover, the chirality of the charge carriers is one of the main ingredients required to have non-reciprocal devices and chiral motion can be achieved in several different ways, through different physical mechanisms.
In the present work we focused only on the quantum Hall effect, but we believe a similar device model can be developed also for other kinds of material, e.g. topological insulators or conductors with finite Berry flux (anomalous Hall effect) \cite{Rudner}.
Further quantitative studies in these situations are now required  to determine the most efficient way to achieve non-reciprocity and to offer convenient alternatives for constructing new low-temperature quantum technologies.

\section{Acknowledgements}
The authors would like to thank D. Reilly, A. Mahoney,
A.C. Doherty, G. Verbiest, C. Stampfer, F. Haput and F. Hassler for useful discussions.
This work was supported by the Alexander von Humboldt foundation.

\begin{appendix}
\section{ \label{app:Eigen-funct}Sharp edges, static eigensystem}

Here, we solve the static, non-interacting Schr{\"o}dinger's equation for 2DEGs terminated by a sharp edge in position $x=0$. 
In the Landau gauge and exploiting the translational invariance in the $y$-direction, the ladder operators in Eq. (\ref{eq:ladder-op-2DEG}) reduce to
\begin{subequations}
\begin{flalign}
\hat{a}^{\dagger} & = \frac{1}{\sqrt{2}}\left(-l_B\frac{\partial}{\partial x}+\frac{x}{l_B}+k_y l_B\right),\\
\hat{a} & =\frac{1}{\sqrt{2}}\left(l_B\frac{\partial}{\partial x}+\frac{x}{l_B}+k_y l_B\right).
\end{flalign}
\end{subequations}

The 2DEG Hamiltonian in Eq.(\ref{eq:MF-ham-2deg}) then becomes
\begin{equation}
\frac{\hat{H}_B}{\hbar \omega_c}=-\frac{l_B^2}{2} \frac{\partial^2}{\partial x^2}+\frac{1}{2}\left(\frac{x}{l_B}+k_y l_B\right)^2,
\end{equation}
and the Schr{\"o}dinger's equation has a general set of normalizable solutions for $x\geq 0$
\begin{equation}
\psi(x)=C e^{-(x/l_B+k_y l_B)^2/2} H_{\epsilon/(\hbar \omega_c)-1/2}(x/l_B+k_y l_B).
\end{equation}
Here, $H_{\alpha}(z)$ is the Hermite function, $\epsilon/(\hbar \omega_c)-1/2$ is a real and positive number and $C$ is the normalization constant.
The boundary condition of vanishing wavefunction at $x=0$, implies the dispersion relation
\begin{equation}
H_{\epsilon/(\hbar \omega_c)-1/2}(k_y l_B)=0.
\label{eq:hermite-dispersion}
\end{equation}
For an analytic approximation of the zeroes of the Hermite functions, see \cite{Hermite}.
Note that in the limit $k_y l_B\ll -1$, one, as expected, obtains the Landau levels in Eq. (\ref{eq:MF-ham-eigenvalue-2deg}), and the corresponding shifted harmonic oscillators eigenfunctions in Eq. (\ref{eq:hermite pol}).

The Hamiltonian for a monolayer graphene, Eq. (\ref{eq:MF-ham-graph}), can be diagonalized in a similar way, but carefully accounting for the additional degrees of freedom. For rather general smooth boundaries, the problem was solved analytically in terms of Hermite functions in \cite{Akhmerov2, Akhmerov4}.

\section{ \label{app:Fringing}Electrostatic Green's function}
Consider a voltage $V_e(\overline{r},t)$ applied to a top-gate at distance $d$ with respect to a 2-dimensional material, with a charge density $\rho(\overline{r},t)$. Here, $\overline{r}\equiv(x,y)^T$.
Neglecting retardation, the problem is purely electrostatic and it can be solved by introducing the electrostatic Green's function and inverting the Poisson equation. In the region between the 2-dimensional material and the electrode, the potential is
\begin{multline}
\phi(\overline{r},z,t)=-\int_{\mathbb{R}^2} d\overline{r}'G(\overline{r},\overline{r}',z)\rho(\overline{r}',t)+ \\
\int_{\mathbb{R}^2} d\overline{r}'G_f(\overline{r},\overline{r}',z)V_e(\overline{r}',t),
\label{eq:Greens-top-gate}
\end{multline}
with
\begin{multline}
G=\frac{1}{4\pi\epsilon_S}\left( \frac{1}{\sqrt{(x-x')^2+(y-y')^2+z^2}} -\right. \\
\left.\frac{1}{\sqrt{(x-x')^2+(y-y')^2+(z-2d)^2}} \right),
\label{eq:Greens-top-gate-3d}
\end{multline}
and
\begin{equation}
G_f(\overline{r},\overline{r}',z)=\frac{1}{4\pi^2}\int_{\mathbb{R}^2} d\overline{q}e^{-i\overline{q}.\overline{r}}e^{q (z-d)}.
\label{eq:Greens-top-gate-ff}
\end{equation}
Here, $\overline{q}\equiv(q_x,q_y)^T$, $q\equiv\sqrt{q_x^2+q_y^2}$, and $G_f$ is determined by the 2-dimensional inverse Fourier transform of an exponential.

$G$ in the first term on the right hand side of Eq. (\ref{eq:Greens-top-gate}) represents the Green's function of a system with a grounded electrode, described by an image charge in position $z=2d$ \cite{Jackson}; the second term fixes the potential of the electrode to $V_e(\overline{r},t)$, also including fringing fields.

Focusing on the potential on the plane of the 2-dimensional electron gas, $z=0$, one can evaluate Eq. (\ref{eq:Greens-top-gate-ff}) by performing an Hankel transform \cite{Bateman}, leading to, for $d\neq0$,
\begin{equation}
G_f(\overline{r},\overline{r}',0)=\frac{1}{2\pi}\frac{d}{\left((x-x')^2+(y-y')^2+d^2\right)^{3/2}}.
\label{eq:Greens-top-gate-ff1}
\end{equation}

For example, for an applied potential of the form $V_e=\Theta(-y)V(t)$, the second term in the right hand side of Eq. (\ref{eq:Greens-top-gate}) becomes
\begin{equation}
\int_{\mathbb{R}^2} d\overline{r}'G_f(\overline{r},\overline{r}',0)V_e(\overline{r}',t)=V(t)\left(\frac{1}{2}-\frac{1}{\pi}\tan^{-1}\frac{y}{d}\right).
\end{equation}

If $d$ is sufficiently small (compared to the length of the electrodes and of the gaps between them), the fringing fields can be neglected and we approximate
\begin{equation}
\int_{\mathbb{R}^2} d\overline{r}'G_f(\overline{r},\overline{r}',0)V_e(\overline{r}',t)\approx V_e(\overline{r},t).
\label{eq:fringing-appr}
\end{equation}

\section{ \label{app:motion-eq}General equation of motion}
Here, we derive Eq. (\ref{eq:motion-eq-general}).
First, we consider that for small energy excitation,  $\omega/ \omega_c \ll 1$, with small momentum transfer, $q_y \mathrm{max}(l_B, w) \ll 1 $, and at low enough temperature, $k_B T/(\hbar \omega_c)\ll 1$, one can neglect mixing of LLs with different quantum number.
Using Eqs. (\ref{eq:matrix-decomp-rho}), (\ref{eq:matrix-element-rho}) and (\ref{eq:Psi-factorization}),  and Fourier transforming the $y$ coordinate, $y\rightarrow q_y$, we get
\begin{multline}
\rho_1=\frac{e}{L_y}\sum_{n,k_y,k'_y} \frac{f_F (\epsilon_{n k'_y})-f_F (\epsilon_{n k_y})}{\epsilon_{n k'_y}-\epsilon_{n k_y} +\hbar (\omega + i \eta) }\times \\
 U_{n k'_y k_y}(\omega) \psi^*_{n k'_y}(x)\psi_{n k_y}(x) \delta(k_y+q_y-k'_y).
\end{multline}

We now take the thermodynamic limit, $L_y\rightarrow\infty$ and promote the continuous momentum quantum numbers to arguments of the functions. The two summations over the momentum become two integrals re-scaled by a factor $L_y/(2\pi)$ each. 
We assume the excitations to be smooth in the $y$-direction, $q_y \mathrm{max} (l_B, w) \ll 1$, and we linearize in $q_y$, leading to
\begin{multline}
\rho_1=\frac{e L_y}{4\pi^2 \hbar }\sum_{n}\int_{\mathbb{R}}dk_y \frac{\frac{\partial f_F \left(\epsilon_n(k_y)\right)}{\partial k_y} q_y}{\frac{\partial \epsilon_n(k_y)}{\hbar\partial k_y} q_y+ \omega + i \eta  }\times \\
 U_{n}(k_y, q_y,\omega)\left| \psi_{n}(x,k_y)\right|^2,
 \label{eq:TDlimit-linearqy}
\end{multline}
where $U_{n}(k_y, q_y,\omega)$ is given from Eqs. (\ref{eq:us-matrix-full}) and (\ref{eq:Psi-factorization}), and in these limits reduces to
\begin{multline}
 U_{n}(k_y, q_y,\omega)=\frac{2\pi}{L_y}\int_{\mathbb{R}}dx'  U(x',q_y,\omega)\left| \psi_{n}(x',k_y)\right|^2.
 \label{eq:us-matrix-app}
\end{multline}
Note that the $L_y$ factors in Eqs. (\ref{eq:TDlimit-linearqy}) and (\ref{eq:us-matrix-app}) cancel out.

Introducing the quantity $p$ and the quantum velocities $v^q$, defined in (\ref{eq:pn-def}) and  (\ref{eq:quantum-velocity}) respectively, we get from Eq. (\ref{eq:TDlimit-linearqy})
\begin{equation}
p_n(k_y,q_y,\omega)=\frac{e L_y}{4\pi^2 \hbar } \frac{q_y  U_{n}(k_y, q_y,\omega)}{v^q_n(k_y) q_y+ \omega + i \eta}.
 \label{eq:pn-usn}
\end{equation}

Finally, to get the self-consistent equation of motion (\ref{eq:motion-eq-general}) for $p$, we combine the explicit form of the screened potential, in Eqs. (\ref{eq:screened-potential}) and (\ref{eq:inverted-Poisson-equation}), with Eqs. (\ref{eq:pn-def}), (\ref{eq:us-matrix-app}) and (\ref{eq:pn-usn}), and we introduce the external potential matrix elements and the electrostatic velocity, defined by (\ref{eq:Ve-matrix}) and (\ref{eq:classical-velocity}) respectively.


\section{\label{app:smooth-edge}Smooth edge limit}
Here, we derive the equation of motion (\ref{eq:motion-smooth-edges}) in the limit of smooth edges. 
We neglect the quantum velocities and all the details at length scales $l_B$, and we approximate
\begin{equation}
\left| \psi_n(x,k_y)\right|^2\approx \delta(x-X),
\end{equation}
where $X\equiv-k_y l_B^2$ is intended to be the center of mass of the electron's wavefunctions.

With these assumptions, the potential matrix elements $V_{n}$ and $U_{n}$, the rescaled charge $p_n$ and the electrostatic velocities $v^c_{n m}$ become independent of the LL quantum numbers and Eqs. (\ref{eq:Ve-matrix}), (\ref{eq:us-matrix-app}), (\ref{eq:pn-usn}) and (\ref{eq:classical-velocity}) reduce to, respectively,
\begin{subequations}
\begin{flalign}
V_{n}\left(k_y,q_y,\omega \right) & \approx V_e(X,q_y,\omega),
\label{eq:ve-smooth}\\
 U_{n}(k_y, q_y,\omega) &\approx \frac{2\pi}{L_y} U(X,q_y,\omega),\\
 p_n(k_y,q_y,\omega)&\approx p(X,q_y,\omega)=\frac{e q_y}{2\pi \hbar } \frac{ U(X, q_y,\omega)}{\omega + i \eta},
 \label{eq:pn-usn-smooth}\\
 v^c_{n m}(k_y,k'_y,q_y)&\approx \frac{e^2}{\hbar} G_0(X,X',q_y).\label{eq:vc-smooth}
\end{flalign}
\end{subequations}
 
Substituting Eqs. (\ref{eq:ve-smooth}), (\ref{eq:pn-usn-smooth}) and (\ref{eq:vc-smooth}) into Eq. (\ref{eq:motion-eq-general}), one gets
\begin{multline}
\left(\omega+ i \eta\right) p\left(X, q_y,\omega \right)=-\frac{e^2 q_y}{2 \pi \hbar } V_{e}\left(X,q_y,\omega \right) +\\
\frac{e^2 q_y}{\hbar }  \int _\mathbb{R} dX' G_0(X,X',q_y)
p\left(X', q_y,\omega \right)\sum _m \frac{\partial f_F\left(\epsilon_m(X')\right)}{\partial X'}.
\label{eq:rho1-smooth1}
\end{multline}

To proceed further and connect to the classical result, we need to introduce the static charge density. In our model, it is given by 
\begin{equation}
\begin{split}
\rho_0(x)&=\frac{1}{2 \pi l_B^2}\sum_n\int_{\mathbb{R}}dX f_F(\epsilon_n(X))\left| \psi_n(x,X)\right|^2\\
	&\approx \frac{1}{2 \pi l_B^2}\sum_n f_F(\epsilon_n(x)),
	\label{eq:rho0-smooth}
\end{split}
\end{equation}
where the prefactor $1/(2\pi l_B^2)$ coincides with the standard density of states of a LL \citep{QuantumHallGirvin}, and the summation of the probabilities corresponds to the local filling factor $\nu(x)$, ranging from zero to the bulk filling factor $\nu_0$.
Note that $\nu(x)$ can have the form described in \citep{CSG} because of the quasi-degeneracy in momentum $k_y$ at the Fermi energy.

Combining Eqs. (\ref{eq:pn-def}), (\ref{eq:pn-usn-smooth}) and (\ref{eq:rho0-smooth}), the linearized charge density can be expressed as
 \begin{equation}
\rho_1(x, q_y, \omega) =2\pi l_B^2 p\left(x, q_y, \omega\right) \frac{\partial\rho_0(x)}{\partial x}.
\label{eq:linearized-charge}
\end{equation}

Identifying now the position $X$ of the center of mass of the infinitely narrow wavefunctions with the $x$-coordinate, substiting Eq. (\ref{eq:linearized-charge}) into Eq. (\ref{eq:rho1-smooth1}), and using the definition of magnetic length $l_B\equiv \sqrt{\hbar/(e |B|)}$ and of cyclotron frequency $\omega_c\equiv eB/m$, we obtain Eq. (\ref{eq:motion-smooth-edges}).

\section{\label{app:admittance}Admittance}

Here, we derive the admittance matrix for the 3-terminal QH gyrator in Eq. (\ref{eq:admittance}).
When the gaps between electrodes are small compared to their length, one can neglect the change of Green's function in these region and use the Green's function in Eq. (\ref{eq:Greens-top-gate-3d}).
Also, we assume that the electrodes are rectangular and they completely cover the plasmon charge distribution in the $x$-direction (normal to the boundary). This allows the decomposition of the surface integral in Eq. (\ref{eq:current-displacement}) into
\begin{equation}
\int_{S_n}d\overline{r}\approx \int_{\mathbb{R}}dx\int_{y_n}^{y_n+L_n}dy.
\end{equation}

Combining Eqs. (\ref{eq:plasmon-full-sum}), (\ref{eq:motion-smooth-eigen-aj}), (\ref{eq:current-displacement}) and (\ref{eq:Greens-top-gate-3d}), one gets, after some straightforward algebra,
\begin{multline}
I_n(t)= \sum_j \int_{\mathbb{R}^2}dx'dy'g_j(x')\left(v_j u_j(y',t)+a_jV_e(y',t)\right)\times \\
\left( \mathcal{L}_{2d}(y'-y_n)-\mathcal{L}_{2d}(y'-y_n-L_n)\right),
\label{eq:I-full-deriv}
\end{multline}
with $\mathcal{L}_\gamma$ being a normalized Lorentzian distribution with standard deviation $\gamma$.
We have neglected the damping rate $\eta$ here: it can be easily incorporated at the end of the calculations by the substitution $\omega\rightarrow \omega+i \eta$.

To proceed further, we impose the condition $2d/L_n\ll 1$, such that the two Lorentzian functions are spatially separated, and we approximate them with Dirac deltas. This approximation is the same as the one used in Eq. (\ref{eq:fringing-appr}), where we neglect the fringing fields at the termination of electrodes.
In this approximation, using an external driving voltage of the form given in Eq. (\ref{eq:voltage-external}), the term proportional to $V_e$ in Eq. (\ref{eq:I-full-deriv}) vanishes and the current is proportional to the difference of the excess charge at position $y_n$ and $y_n+L_n$.

Fourier trasforming in time, $t\rightarrow \omega$, Eq. (\ref{eq:motion-smooth-eigen-aj}), and using Eq. (\ref{eq:voltage-external}), the EMP charge can be written as
\begin{equation}
u_j(y,\omega)=\sum_{m=1}^N  \frac{a_j}{2 v_j} \upsilon_j^m(y,\omega) V_m(\omega), 
\end{equation}
with
\begin{equation}
\begin{split}
\upsilon_j^m(y,\omega)= & -\left( 1+i \cot\left(\frac{\omega L_y}{2 v_j}\right)  \right) \times \\
& \left( \Theta(y-y_m)e^{i \omega(y-y_m)/v_j}- \right. \\
 & \left. \Theta(y-y_m-L_m)e^{i \omega(y-y_m-L_m)/v_j}+ \right. \\
 & \left. \Theta(y_m-y)e^{i \omega(y+L_y-y_m)/v_j}- \right. \\ 
 & \left. \Theta(y_m+L_m-y)e^{i \omega(y+L_y-y_m-L_m)/v_j} \right) .
\end{split}
\label{eq:charge-fourier}
\end{equation}

Combining Eqs. (\ref{eq:I-full-deriv}) and (\ref{eq:charge-fourier}) in the Dirac delta approximation, and introducing 
\begin{equation}
q_j=\frac{ a_j}{2} \int_{\mathbb{R}}dx' g_j(x'),
\end{equation}
we can decompose the current flowing in each electrode of  a $N$-terminal device as
\begin{equation}
I_n=\sum_{m=1}^N Y_{nm}(\omega) V_m(\omega),
\label{eq:general-current}
\end{equation}
with general admittance matrix element 
\begin{equation}
Y_{n m}(\omega)=\sum_{j}q_j \left( \upsilon_j^m(y_n, \omega)-\upsilon_j^m(y_n+L_n, \omega) \right).
\label{eq:general-admittance-element}
\end{equation}

Explicitly calculating $q_j$ for smooth and sharp edges and simplifying Eqs. (\ref{eq:general-current}) and (\ref{eq:general-admittance-element}) for the 3-terminal device in Fig. \ref{fig:3-terminals-gyrator}, we obtain Eq. (\ref{eq:admittance}).

\end{appendix}

\bibliography{lit}

\end{document}